\documentclass[11pt,twoside]{article}
\usepackage{epsfig}
\usepackage{here}
\usepackage{amsmath}
\usepackage{amssymb}
\usepackage{bbm}
\newcommand{\beq}{\begin{equation}}
\newcommand{\eeq}{\end{equation}}
\newcommand{\beqa}{\begin{eqnarray}}
\newcommand{\eeqa}{\end{eqnarray}}
\begin{document}
%
%
%

\begin{center}{\Large\bf ISI and FSI in $NN\rightarrow NNX$ reactions\\[2mm]
close to threshold}
\end{center}
\vspace{0.5cm}
\begin{center}
     F.\ Kleefeld \footnote{e-mail: {\sf kleefeld@cfif.ist.utl.pt}, 
 URL: {\sf http://cfif.ist.utl.pt/$\sim$kleefeld/}}${}^{,}$\footnote{Invited contribution to the symposium on {\em Threshold meson production in $pp$ and $pd$ interaction}, extended \mbox{COSY-11} collaboration meeting, 20.-24.6.2001, Cracow, Poland.} \\[0.1cm]
     {\small \em Centro de F\'{\i}sica das Interac\c{c}\~{o}es Fundamentais, 
Instituto Superior T\'{e}cnico,} \\
{\small \em Edif\'{\i}cio Ci\^{e}ncia, Piso 3,  
Av. Rovisco Pais, 
P-1049-001 LISBOA,
Portugal}
\end{center}
\vspace{0.5cm}
\begin{center}
 \parbox{0.9\textwidth}{
  \small{
   {\bf Abstract:}\ The present status of the theoretical description of Initial (ISI) and Final State Interactions (FSI) in nucleon induced exclusive meson production processes close to threshold is reviewed. Shortcomings are addressed and new perspectives given.}}
\end{center}

\vspace{0.5cm}
\section{Motivation}
It is quite surprising, that the discussion of the theoretical description of ISI and FSI in nuclear and particle collision/production processes is gaining again increasing attention, even though people thought, this issue has been well understood due to an established formalism going back to outstanding scientists such as e.g.\ K.\ Brueckner, G.\ Chew, E.\ Hart, K.M.\ Watson, A.B.\ Migdal, E.\ Fermi, G.\ Gamov and A.\ Sommerfeld entering nearly all standard text books of Scattering Theory, e.g. \cite{gol64,joa83}. The existing formalism for the treatment of ISI and FSI is based on the idea, that T-matrix elements of many processes (in the Distorted Wave Born Approximation (DWBA)) are well described as overlap integrals of initial and final state wavefunctions multiplied by some interaction potential. This idea works quite successfully for (mainly nonrelativistic) textbook examples, in which excitation energies are small compared to the masses of the particles involved, i.e. for the description of electromagnetic transitions in atoms (excitation energies $\sim 1$ eV) or $\beta^\pm$-decays/transitions of atomic nuclei (excitation energies $\sim$ \mbox{10${}^2\ldots$10${}^3$ keV}). In these processes the initial/final states hardly go offshell due to the transition interaction, or at least for a very short time.
In reaction processes like $NN\rightarrow NNX$ (with $X$ being a mesonic system) close to the particle production threshold the situation is quite different. The excitation energy of the $NN$ system in such a process is of the order of the mass of $X$. In the case of $\eta$-production this yields already about half of the nucleon mass, for heavier mesons the excitation energy is in the range of the nucleon mass. At high excitation energies is a difficult task for the outgoing $NNX$ system to rearrange  the 3-momenta and bring the 3 outgoing particles to their mass-shells. In the case of an incoming or outgoing $pp$ system the infinite range of the Coulomb force even prevents the initial or final state particles respectively to go to their mass shell within a finite time. The production of ``light'' $\pi$-mesons seems to be quite exceptional. Yet even there one can see already various new aspects being important for the description of ISI and FSI of processes under consideration.
The issue of this presentation is not the discussion of various experimental signals for mainly FSI in $NN\rightarrow NNX$ shown e.g.\ in \cite{bil98,mos00b,mos00} (and references therein). It is a introductory review of present theoretical formalisms and new theoretical developments treating ISI/FSI to enable researchers  involved to find to a (still outstanding) {\em quantitative} covariant theoretical formulation to describe ISI/FSI and a better quantitative analysis of experimental data.
\section{Invariants and notation in $NN\rightarrow NNX$}
In the following I will introduce quantities used in the theoretical description of the reaction $NN\rightarrow NNX$. For convenience I use the metric $(+,-,-,-)$, denote the outgoing meson $X$ by $\phi$, the incoming/outgoing particles by \mbox{$1$, $2$, $1^\prime$, $2^\prime$ and $3^\prime$}, i.e.\ $N(1)N(2)\rightarrow N(1^\prime)N(2^\prime)\,\phi (3^\prime)$, and the  respective 4-momenta by $p_1$, $p_2$, $p_{1^\prime}$, $p_{2^\prime}$ and  $p_{3^\prime} = k_\phi$. 5 independent Lorentz invariants $s$, $s_1$, $s_2$, $t_1$, $t_2$ and 1 dependent invariant $s_3$ are introduced by $s := (p_1+p_2)^2$, $s_1 := (p_{1^\prime}+p_{2^\prime})^2$, $s_2 := (p_{2^\prime}+p_{3^\prime})^2$, $t_1 := (p_1-p_{1^\prime})^2$, $t_2 := (p_2-p_{3^\prime})^2$ and $s_3 := (p_{3^\prime}+p_{1^\prime})^2 = m^2_{1^\prime}+m^2_{2^\prime}+m^2_{\phi}-\,s_1-\,s_2$.
For a discussion of the 3-body phasespace I refer the reader e.g.\ to \cite{byc73,alv73,pie83}. Using the kinematical triangle function $\lambda (x,y,z) := x^2+y^2+z^2 - 2\, (xy + yz + zx)$ and defining $|\vec{k}_\phi|_{cm} := \sqrt{\lambda (s,m^2_\phi,s_1) /(4\,s)}$ there are various ways to introduce quantities, which carry the same energy information as the cm-energy $\sqrt{s}$, e.g. $Q_{cm} := \sqrt{s} - \, m_{1^\prime} - \, m_{2^\prime} - \, m_\phi$, $|\vec{p}_1|_{cm} = |\vec{p}_2|_{cm} := \sqrt{\lambda (s,m^2_1,m^2_2)/(4\,s)}$, $\kappa_0 := \sqrt{\lambda (s,m^2_1,m^2_2)}/(2\,(m_1+m_2))$ and $\eta_\phi := |\vec{k}_\phi|^{max}_{cm}/m_\phi = \sqrt{\lambda (s,m^2_\phi,s^{\,min}_1)/(4\,s \, m^2_\phi)}$
with $s^{\,min}_1 = (m_{1^\prime}+m_{2^\prime})^2$. Here $\kappa_0$ is the Lab-wavenumber of the relative motion of the incoming $NN$-system. The {\em excess-energy} $Q_{cm}$ and the dimensionless quantity $\eta_\phi$, which is the maximum possible cm-momentum of the produced meson in units of the meson mass $m_\phi$, vanish exactly at threshold. The relative motion of the outgoing 2-particle subsystems $1^\prime 2^\prime$, $1^\prime 3^\prime$ and $2^\prime 3^\prime$ can be described e.g.\ in terms of relative Lab-wavenumbers $\kappa_{23}$, $\kappa_{31}$ and $\kappa := \kappa_{12}$:
\begin{equation} \kappa_{12}:= \frac{\sqrt{\lambda (s_1,m^2_{1^\prime},m^2_{2^\prime})}}{2\,(m_{1^\prime}+m_{2^\prime})} , 
\, \kappa_{23} := \frac{\sqrt{\lambda (s_2,m^2_{2^\prime},m^2_\phi)}}{2\,(m_{2^\prime}+m_\phi)} , \, \kappa_{31} := \frac{\sqrt{\lambda (s_3,m^2_{1^\prime},m^2_\phi)}}{2\,(m_{1^\prime}+m_\phi)}  
\, . 
\end{equation}
\section{The Watson-Migdal Approach} \label{sec3}
\subsection*{Historical remark}
The present theoretical formalism of treating ISI and FSI goes mainly back to ideas formulated by K.\ Brueckner, G.\ Chew and E.\ Hart (1951) \cite{bru51}, K.M.\ Watson (1951,1952) \cite{wat51,wat52}, A.B.\ Migdal (1955) \cite{mig55} and E.\ Fermi (1955) \cite{fer55}, and has been transported through the literature by now as {\em Watson-Migdal \mbox{Approximation}} (WMA) or {\em Distorted Wave Born Approximation} (DWBA). The idea behind this formalism is, that the T-matrix $T_{fi}$ of a particle production/anihilation/scattering process can be ``factorised'' into bound-state or (elastic) scattering wavefunctions describing the particles in the initial and/or final state, and a short-ranged interaction part $T^{\,(0)}$, describing the transition from the initial to the final state. The T-matrix in the Watson-Migdal Approach can therefore be factorised in dimensionless {\em enhancement factors} $T_{FSI}$ and $T_{ISI}$ describing the influence of FSI and ISI respectively (being 1 for no FSI and ISI respectively) and a short-ranged T-matrix $T^{\,(0)}$ describing the production/scattering process without ISI/FSI:
\begin{equation} T_{fi} \; \simeq \; T_{FSI} \; T^{\,(0)} \: T_{ISI} \, .\label{fac2} \end{equation}
In Watson's first extensive approach \cite{wat52} considering a process $12\rightarrow 1^\prime 2^\prime$ the energy dependence of the T-matrix and therefore of the (differential) cross section $d\sigma/d\kappa$ stems mainly from the 2-particle final state wavefunction  $\psi_{f,L}$, which is asymptotically related to the wavenumber $\kappa$, phaseshifts $\delta_L$ and the scattering length $a_L$ ($L$ = orbital angular momentum of the $1^\prime 2^\prime$ system)\footnote{K.\ Brueckner et al.\ \cite{bru51} developed the idea, that $T_{FSI}$ is given by the wavefunctions at the interaction point $r=0$, i.e. by $\psi_f(0)$. This idea has been later used by R.\ van Royen and V.F.\ Weisskopf \cite{roy67} in the initial state to estimate decay rates $\Gamma$ by $\Gamma \propto |\psi_i(0)|^2$.}:
\begin{equation} T_{fi} \propto \psi_{f,L} \sim \frac{e^{\,i\,\delta_L} \sin \delta_L}{\kappa^{\,L + 1}} = \frac{\kappa^{\,L}}{\kappa^{\,2\,L + 1} (\cot \delta_L - i)} \simeq - \, \frac{a_L \, \kappa^L}{1 + i \, a_L \, \kappa^{\,2\,L + 1} } \; . 
\end{equation}
Here I used the {\em Effective Range Expansion} (ERE) $\kappa^{\,2\,L + 1} \cot\delta_L = -\, a^{-1}_L + O (\kappa^2)$ \cite{che49}. As a consequence Watson obtained for the square of the T-matrix and the differential cross section the following wavenumber dependences:
\begin{equation} |T_{fi}|^2 \propto \frac{|a_L|^2 \, \kappa^{\,2\,L}}{1+ |a_L|^2 \, \kappa^{\,4\,L + 2}} \, \Rightarrow \, \frac{d\sigma}{d\kappa} \propto \frac{\sin^2 \delta_L}{\kappa^{\,2\,L}} \propto \kappa^2 \, |T_{fi}|^2 \propto \frac{|a_L|^2 \, \kappa^{\,2\,L + 2}}{1+ |a_L|^2 \, \kappa^{\,4\,L + 2}}\; .
\end{equation}
Watson's arguments have been generalised to processes with 3 particles in the final state. Let's e.g.\ consider the process $NN\rightarrow NN\phi$ close to threshold. Assuming for simplicity that FSI are mainly described by the outgoing destorted $NN$ wavefunction $\psi_{f,L}$, while representing the produced meson $\phi$ by a spherical wave, which is for the orbital angular momentum $\ell$ of the meson relative to the $NN$ system just proportional to a spherical Bessel function $j_\ell (q^{\,\prime} \, r) \stackrel{q^{\,\prime}\rightarrow \,0}{\propto} q^{\,\prime\,\ell}$ (with $q^{\,\prime} := |\vec{k}_\phi|_{cm}$), one could --- according to Watson --- suggest $T_{fi} \propto q^{\,\prime\,\ell} \, \psi_{f,L}$. Including the 3-body phasespace one obtains {\em within a nonrelativistic framework} for the total cross section:
\begin{equation} \sigma \; \propto \; \int^{\, m_\phi \, \eta_\phi}_0 dq^\prime \;q^{\,\prime\,2} \, \kappa \; |T_{fi}|^2 \; \propto \; \int^{\, m_\phi \, \eta_\phi}_0 dq^\prime \;q^{\,\prime\;2\,\ell + 2} \; \frac{|a_L|^2 \, \kappa^{\,2\,L + 1}}{1+ |a_L|^2 \, \kappa^{\,4\,L + 2}}\; . \label{eqx2}
\end{equation}
Using the {\em threshold} identity $\kappa \stackrel{threshold}{\longrightarrow} \sqrt{(m^2_\phi \,\eta^2_\phi - q^{\,\prime\,2}) \left(\frac{m_\phi + 2\, m_N}{4\,m_\phi} + O(q^{\,\prime\,2})\right)}$ a Taylor expansion in $\eta_\phi$ yields for the cross section close to threshold:
\begin{equation} \sigma\, \Big(\, NN\rightarrow \,[ [ NN\, ]_L \,\phi \,]_{\ell}\,\Big) \; \propto \; \eta_\phi^{\,2\,(\ell + L) \,+ \,4} \; . \label{eqx1}
\end{equation}  
\subsection*{Derivation of the Onshell-Watson-Migdal Approximation}
Now I am going to derive a more sophisticated Watson-Migdal formalism, which yields the traditional Watson-Migdal Approach in a certain limit\footnote{The formalism to be presented is essentially equivalent to the so called {\em two potential scattering formalism} which can be found in various textbooks of quantum scattering theory (see e.g.\ \cite{joa83}). Throughout such a formalism the potential consists of a sum of a short- and long-ranged potential.}. The derivation --- already fully performed by Watson --- has recently been rederived and critically reconsidered by the author in \cite{kle99a,kle99c,kle00b}, while simultaneously promoted in \cite{han99}. 
Let's consider a system of colliding and produced particles being described by a 
Hamilton operator $H = K + V = K + W + U$, with $K$ being the kinetic energy
operator, $W$ being the particle number nonconserving short range interaction
potential and $U$ being the particle number conserving interaction potential of long range. Furthermore I define the long range Hamilton operator
$h=K+U$. The respective eigenstates to the operators $H$, $h$ and $K$ fulfil the following Schr\"odinger equations:
\begin{equation}
(E-H)|\psi^\pm_\alpha > \; = \; 0 \; , \quad
(E-h)|\chi^\pm_\alpha > \; = \; 0 \; , \quad
(E_n-K)|\varphi_n > \; = \; 0 
\end{equation}
yielding the Lippmann--Schwinger equations ($G_0^\pm (E,K):=(E-K\pm i\varepsilon)^{-1}$):
\begin{eqnarray}
|\psi^\pm_\alpha > \, = & \displaystyle
|\varphi_\alpha > + \frac{1}{E-H\pm i\varepsilon}\, V |\varphi_\alpha >
\, = & |\chi^\pm_\alpha > + \,\frac{1}{E-H\pm i\varepsilon}\, W  |\chi^\pm_\alpha > \nonumber \\
|\chi^\pm_\alpha > \,\,= & \displaystyle
|\varphi_\alpha > + \frac{1}{E-h\pm i\varepsilon}\, U |\varphi_\alpha >
\,\, = & |\varphi_\alpha > + \,G_0^\pm (E,K) T^\pm_{el}(E,K) |\varphi_\alpha
> . \nonumber 
\end{eqnarray}
The S-matrix elements of the considered process are determined by 
$S_{\beta\alpha} = \;\;<\psi^{-}_\beta |\psi^{+}_\alpha >$ 
yielding the T-matrix
$T_{\beta\alpha} \; = \; <\psi^{-}_\beta |(U+W)|\psi^{+}_\alpha >$. For $E=E_i=E_f$ the T-matrix can be reformulated in terms of {\em Watson's Theorem} (see e.g.\ p.\ 448, 461 in \cite{joa83}), i.e.: 
\begin{equation} 
T_{\beta\alpha} \; = \; <\chi^{-}_\beta |U|\varphi_\alpha > \; + \;
<\chi^{-}_\beta |W| \psi^{+}_\alpha > \; = \; < \varphi_\beta |U| \chi^{+}_\alpha > \; + \;
< \psi^{-}_\beta |W| \chi^{+}_\alpha > 
\end{equation}
As the particle number conserving potential $U$ is positioned in matrix elements of states with different particle number, the respective matrix elements have to vanish. As a result I obtain
$ T_{\beta\alpha} \; = \;
<\chi^{-}_\beta |W| \psi^{+}_\alpha > \; = \; 
< \psi^{-}_\beta |W| \chi^{+}_\alpha >$.
In the following step we perform the first approximation widely known as {\em Watson-Migdal Approximation} (WMA)\footnote{This approximation yields a ``factorisable'' T-matrix for a field-theory, which is by construction known to be not factorisable. To the knowledge of the author there is no theoretical work to quantify the validity of this approximation. Even the consequences of the approximation to the unitarity of the approximated S-matrix for systems of many coupled channels have to be investigated.}. The exact asymptotic states $| \psi^{\pm}_\alpha >$ are approximated by the $U$-destorted asymptotic states $| \chi^{\pm}_\alpha >$, i.e.\ $| \psi^{\pm}_\alpha > \simeq | \chi^{\pm}_\alpha >$. As a consequence I obtain an approximated Watson's theorem $T_{\beta\alpha} \simeq <\chi^{-}_\beta |W| \chi^{+}_\alpha >$. This approximation is valid to first order in $W$ and to all orders in $U$.
Defining the free propagator $G_0^\pm (E,K)= (E-K\pm i\varepsilon)^{-1}$ and inserting complete sets of free asymptotic states $|\phi_n>$ the T-matrix in the WMA can be separated into a short- and long-ranged part via:
\begin{eqnarray}
T_{\beta\alpha} & \simeq &
 <\varphi_\beta|\Big(1+\sum\limits_n T^+_{el}(E,K)|\varphi_n > G_0^+ (E,E_n) <\varphi_n |\Big) \; W 
\nonumber \\
 & & \Big( 1 + \sum\limits_m |\varphi_m > G_0^+ (E,E_m) <\varphi_m |T^+_{el}(E,K)\Big) |\varphi_\alpha> \, . \label{tmat1}
\end{eqnarray}
If the incoming/outgoing particles are only shortly/weekly offshell, one may approximate the free propagator $G_0^\pm (E,K)$ in the following way\footnote{This approximation I will call here {\em Onshell-Watson-Migdal Approximation} (OWMA).}: 
\begin{equation}
G_0^\pm (E,K) \, = \, \frac{1}{E-K\pm i\varepsilon} \, = \, 
P \,\frac{1}{E-K} \; \mp \; i\pi \delta(E-K) \, \approx \,
\mp \; i\pi \delta(E-K) \, .
\end{equation}
The remaining $\delta$-distribution enforces the incoming or outgoing particles respectively to go to their mass shell and to scatter elastically. This is why I say, the remaining principle value part of the propagator contains all the ``offshell'' information of the scattering/production process,{\em  which is in problems with many coupled channels needed to restore the unitarity of the S-matrix}\footnote{K.\ Nakayama is surely right, if he states, that offshell effects have not to be observable within a field-theoretical formalism. He claims, that results of the principle value integrals are cutoff dependent and therefore dependent on the renormalisation scheme. To the opinion of the author this is due to the approximations made within the Watson-Migdal Approach. A careful parametrisation of the (unknown) principle value contribution of the propagator restoring field-theoretical constraints like unitarity of the S-matrix etc. will again remove the renormalisation-scheme dependence of the Watson-Migdal Approach. The discussion about observability of ``offshell'' effects shows, how important and instructive a careful quantitative formulation of the treatment of ISI and FSI is.}. Let's see, how the OWMA shows up in a non-relativistic treatment of a simple $12\rightarrow \,1^\prime2^\prime$ scattering process, in which both the initial and final state are treated onshell. In order to proceed I identify the free asymptotic states $|\varphi_n>$ with momentum eigenstates $|\vec{\kappa}_i>$ and $|\vec{\kappa}_f>$ in the initial and final state respectively, while in the intermediate state I insert complete sets of momentum eigenstates $|\vec{q}_i>$ and $|\vec{q}_f>$ with $\int d^3 q_i\, |\vec{q}_i><\vec{q}_i|$ $=\int d^3 q_f\,|\vec{q}_f><\vec{q}_f| = \,1$ resembling the symbolic completeness relation $\sum_n |\varphi_n><\varphi_n|=\,1$. Using the nonrelativistic dispersion relation $E=\kappa^{\,2}_f /(2m) = \kappa^{\,2}_i /(2m)$ the T-matrix (\ref{tmat1}) in the OWMA is given explicitly by: 
\begin{eqnarray} \lefteqn{T (\vec{\kappa}_f;\vec{\kappa}_i) \; \stackrel{onshell}{\simeq}}\nonumber \\
 & \simeq &
 \Big(\, <\vec{\kappa}_f| - \, i\,\pi \int d^3 q_f <\vec{\kappa}_f| T^+_{el}(E,K)|\vec{q}_f> \, \delta (E - \frac{|\vec{q}_f|^2}{2\,m})\, <\vec{q}_f|\,\Big)\,W 
\nonumber \\
 & & \Big( \, |\vec{\kappa}_i> - \, i \, \pi \int d^3 q_i \, |\vec{q}_i> \,\delta (E - \frac{|\vec{q}_i|^2}{2\,m}) \, <\vec{q}_i| T^+_{el}(E,K) |\vec{\kappa}_i>\Big) \, .
\end{eqnarray}
Now one can perform a partial wave expansion of the elastic T-matrices in the initial and final state.
The result is given in the footnote\footnote{The axially symmetric partial wave expansions of the elastic T-matrices are given by:
\begin{eqnarray}
\lefteqn{- \,(2\,\pi)^2\,m\; <\vec{\kappa}_f| \, T^+_{el}(E,K)\,|\vec{q}_f> \; =} \nonumber \\
 & & \qquad\qquad = \quad  \frac{1}{\kappa_f} \; \sum\limits^{\infty}_{L_f=\,0} \; (2\, L_f +1)\; e^{i\,\delta^{\,f}_{L_f} (\kappa_f)} \, \sin \delta^{\,f}_{L_f} (\kappa_f) \; P_{L_f} (\cos <\!\!\!)\, (\vec{\kappa}_f,\vec{q}_f) ) \nonumber \\
\lefteqn{- \, (2\,\pi)^2\,m\; <\vec{q}_i| \, T^+_{el}(E,K)\,|\vec{\kappa}_i> \; =} \nonumber \\
 & & \qquad\qquad = \quad \frac{1}{\kappa_i} \; \sum\limits^{\infty}_{L_i=\,0} \; (2\, L_i +1)\; e^{i\,\delta^{\,i}_{L_i} (\kappa_i)} \, \sin \delta^{\,i}_{L_i} (\kappa_i) \; P_{L_i} (\cos <\!\!\!)\, (\vec{q}_i,\vec{\kappa}_i) ) \, . \nonumber 
\end{eqnarray}
Denoting $T (\vec{\kappa}_f;\vec{\kappa}_i) \stackrel{onshell}{\simeq} \sum_{L_f,L_i} \, T (\vec{\kappa}_f,L_f;\vec{\kappa}_i,L_i)\;$ I finally obtain in the OWMA:
\begin{eqnarray} \lefteqn{T (\vec{\kappa}_f,L_f;\vec{\kappa}_i,L_i) \quad =} \nonumber \\
 & = & \Big( 1 + i \, e^{i\,\delta^{\,f}_{L_f} (\kappa_f)} \, \sin \delta^{\,f}_{L_f} (\kappa_f) \Big) \, < \vec{\kappa}_f,L_f| W |\vec{\kappa}_i,L_i> \, \Big( 1 + i \, e^{i\,\delta^{\,i}_{L_i} (\kappa_i)} \, \sin \delta^{\,i}_{L_i} (\kappa_i) \Big) \nonumber \\
 & = & \qquad\qquad\quad\;\; \frac{\cot \delta^{\,f}_{L_f} (\kappa_f)}{\cot \delta^{\,f}_{L_f} (\kappa_f) -i} \; < \vec{\kappa}_f,L_f| W |\vec{\kappa}_i,L_i> \; \frac{\cot \delta^{\,i}_{L_i} (\kappa_i)}{\cot \delta^{\,i}_{L_i} (\kappa_i) -i} \nonumber \\
 & = & \quad\qquad\qquad\qquad\;\; \frac{\mbox{Re} f^{\,f}_{L_f} (\kappa_f)}{f^{\,f}_{L_f} (\kappa_f)} \; < \vec{\kappa}_f,L_f| W |\vec{\kappa}_i,L_i> \; \frac{\mbox{Re} f^{\,i}_{L_i} (\kappa_i)}{f^{\,i}_{L_i} (\kappa_i)} \; .
\end{eqnarray}
Here $f^{\,i}_{L_i} (\kappa_i)$ and $f^{\,f}_{L_f} (\kappa_f)$ denote the well-known {\em Jost functions} (see e.g.\ \cite{joa83}) of the elastic scattering problems in the initial and final state respectively.}. It is obviously of the form (\ref{fac2}), i.e.\ $T_{fi} \; \simeq \; T_{FSI} \; T^{\,(0)} \: T_{ISI}$. It is straight forward to generalise the result to the ``offshell case'', which contains also information about the principle values in the propagators $G_0^\pm (E,K)$. Parametrising the principle value contribution by some unknown function $P(\kappa)$ and denoting $T^{\,(0)} := < \vec{\kappa}_f,L_f| W |\vec{\kappa}_i,L_i>$ the T-matrix in the WMA can be expressed as follows\footnote{The offshell parametrisation $P(\kappa)$ is related to the offshell quantity ${\cal P} (\kappa)$ introduced in \cite{han99} by $P(\kappa) = i \,{\cal P} (\kappa) / (a_L \, \kappa^{2\,L + 1})$ with $\kappa^{2\,L + 1} \, \cot \delta_L (\kappa) = -\, a^{-1}_L + O (\kappa^2)$.}:  
\begin{eqnarray} T (\vec{\kappa}_f,L_f;\vec{\kappa}_i,L_i) & = & T_{FSI} (L_f) \;\; T^{\,(0)} \;\; T_{ISI} (L_i) \quad = \nonumber \\
 & = & \Big( 1 + i \, e^{i\,\delta^{\,f}_{L_f} (\kappa_f) } \, \sin \delta^{\,f}_{L_f} (\kappa_f) \, (1-\,P_f(\kappa_f)) \Big) \, T^{\,(0)} \nonumber \\
 & & \Big( 1 + i \, e^{i\,\delta^{\,i}_{L_i} (\kappa_i)} \, \sin \delta^{\,i}_{L_i} (\kappa_i) \,(1-\,P_i(\kappa_i)) \Big) \nonumber \\
 & = & \Big( 1 + \frac{1}{2} \, ( e^{2 \, i\,\delta^{\,f}_{L_f} (\kappa_f) } \, - 1 ) \, (1-\,P_f(\kappa_f)) \Big) \, T^{\,(0)} \nonumber \\ 
 & & \Big( 1 + \frac{1}{2} \, ( e^{2 \, i\,\delta^{\,i}_{L_i} (\kappa_i) } \, - 1 ) \,(1-\,P_i(\kappa_i)) \Big)  
 \end{eqnarray}
or
\begin{eqnarray} \lefteqn{T (\vec{\kappa}_f,L_f;\vec{\kappa}_i,L_i) \quad = \quad T_{FSI} (L_f) \;\; T^{\,(0)} \;\; T_{ISI} (L_i) \quad =} \nonumber \\
\nonumber \\
 & = & \qquad\qquad\;\; \frac{\cot \delta^{\,f}_{L_f} (\kappa_f)-\,i\, P_f(\kappa_f)}{\cot \delta^{\,f}_{L_f} (\kappa_f)\, -i\qquad\;\;\;} \; T^{\,(0)} \; \frac{\cot \delta^{\,i}_{L_i} (\kappa_i)-\,i\, P_i(\kappa_i)}{\cot \delta^{\,i}_{L_i} (\kappa_i)\, -i\qquad\;\,} \nonumber \\
 & = & \frac{\mbox{Re} f^{\,f}_{L_f} (\kappa_f) + i \, P_f(\kappa_f)\, \mbox{Im } f^{\,f}_{L_f} (\kappa_f)}{f^{\,f}_{L_f} (\kappa_f)} \, T^{\,(0)} \, \frac{\mbox{Re} f^{\,i}_{L_i} (\kappa_i)+ i \, P_i(\kappa_i)\, \mbox{Im } f^{\,i}_{L_i} (\kappa_i)}{f^{\,i}_{L_i} (\kappa_i)}  . \nonumber \\
\label{eqx3} \end{eqnarray}
It is interesting to see 2 ways how to ``switch off'' ISI or FSI: one could perform the limit $\delta_L\rightarrow 0$, or one could set the $P(\kappa) \rightarrow 1$. The OWMA is obtained by $P(\kappa)\rightarrow 0$. In the limit $P(\kappa)\rightarrow -1$ the enhancement factors just reduce to a multiplicative complex phase\footnote{The careful reader may be reminded in the so called {\em Fermi-Watson Theorem} (see e.g.\ \cite{joa83}) based on the idea of the unitarity of the S-matrix, which claims $T_{fi} = \exp (i\,\mbox{Re}\,\delta_f) \, |T_{fi}| \, \exp (i\,\mbox{Re}\,\delta_i)$ and $\mbox{Im}\,\delta_f = \mbox{Im}\,\delta_i$.}, i.e. $T_{ISI} \rightarrow \exp \,(2\,i\,\delta^{\,i}_{L_i} (\kappa_i))$ or $T_{FSI} \rightarrow \exp \,(2\,i\,\delta^{\,f}_{L_f} (\kappa_f))$. It has to be explored in future, in how far the restoration of the unitarity of the S-matrix will set some restrictions on the allowed values of the offshell parametrisation and its limits, and between the offshell parametrisations in the initial and final state\footnote{Among the first few models studying the principle value contribution or offshell effects in ISI and FSI there are the following three to be mentioned: Ch.\ Hanhart and K.\ Nakayama \cite{han99} investigate the ``half-shell'' function related to $P(\kappa)$ within a separable potential model. \mbox{V.\ Baru} et al.\ \cite{bar00,bar01} find essential differences between separable potential models and realistic $NN$ potentials. They calculate $P(\kappa)$ in $NN\rightarrow NNX$ for a simple rescattering graph. It would be very interesting to see, what will change in their results, if the produced meson is coupling to an excited resonance of the nucleon. Finally I want to refer to the work of B.O.\ Kerbikov \cite{ker00}. He compares for the first time in a systematic way the effect of ``half-shell'' and the Jost functions to enhancement factors.}.
$T_{ISI}$ and $T_{FSI}$ derived above are also valid for complex phaseshifts, i.e.\ even in the case of inelasticities yielding $|\eta_L|<1$ for $\eta_L:=\exp(2\,i\, \delta_L) = |\eta_L|\,\exp(2\,i \,\mbox{Re} \, \delta_L)$.
\subsection*{Enhancement factors, 2-potential formalism and unitarity of the S-matrix}
To get a better understanding of the properties of enhancement factors I want to ``define'' them in a very schematic way within the standard 2-potential formalism of Quantum Scattering Theory (see e.g.\ \cite{joa83}). For a 1-channel scattering problem with a potential $V$, which is a sum of a short- and long-ranged potential, i.e.\ $V=V_S+V_L$, the S-matrix $S$ is given in terms of the corresponding phaseshifts by $S =\exp\,(2\,i\,\delta) = \exp\,(2\,i\,(\delta_S+\delta_L) )$, while the T-matrix can be decomposed in the standard manner:
\begin{equation} T \;=\;\frac{e^{\, i\,\delta} \sin\delta}{\kappa} \; = \; \underbrace{\frac{e^{\, i\,\delta_L} \sin\delta_L}{\kappa}}_{=\,T_L} \; + \; e^{\,2\,i\,\delta_L } \; \underbrace{\frac{e^{\, i\,\delta_S} \sin\delta_S}{\kappa}}_{=\,T_S} \; = \; T_L \; + \; e^{\,2\,i\,\delta_L } \; T_S \, .
\end{equation}
Let's now consider a many-channel scattering problem. Here the unitary S-matrix is determined in terms of ``phaseshift matrices'' $\Delta$, $\Delta_S$ and $\Delta_L$ by $S = (S^{-1})^+ =\exp\,(2\,i\,\Delta) = \exp\,(2\,i\,(\Delta_S+\Delta_L) )$ such, that $\Delta = \Delta_S+\Delta_L$ is {\em Hermitian}. The question is now, whether it is possible to define an invertible matrix $T_{FSI} = T^{-1}_{ISI}$ such, that $S =\exp\,(2\,i\,\Delta) = T_{FSI} \; \exp\,(2\,i\,\Delta_S) \; T_{ISI} = (S^{-1})^+$. If $\Delta_S$ is Hermitian (yielding of course also the Hermiticity of $\Delta_L$), the unitarity of the S-matrix $S = (S^{-1})^+$ leads to $T^+_{FSI}=T^{-1}_{FSI}$ and $T^+_{ISI}=T^{-1}_{ISI}$. 
Remembering, that due to the Baker-Campell-Hausdorff formula we can shift the argument $x$ of a function $f$ according to $f(x+a)=\exp\,(a\,\frac{d}{dx}) \, f(x)\, \exp\,(-a\,\frac{\partial}{\partial x})$ by a differential operator $\frac{\partial}{\partial x}$ obeying $[\frac{\partial}{\partial x}\,,\, x]=1$, we can ask the question, whether there exist a ``functional derivative matrix'' $\frac{\delta}{\delta\Delta_S}$ such, that $\left[ \frac{\delta}{\delta\Delta_S}\, ,\, \Delta_S \right] = 1$, yielding a factorisation of the S-matrix $S \; = \; \exp\,\Big(2\,i\,(\Delta_S+\Delta_L)\Big)$  and T-matrix $T = (S-1)/(2\,i\,\kappa )$, i.e.:
\begin{eqnarray}  S & = & \exp \left(\Delta_L \, \frac{\delta}{\delta\Delta_S} \right) \;\exp\,\Big(2\,i\,\Delta_S\Big)\;\exp \left(-\, \Delta_L \, \frac{\delta}{\delta\Delta_S} \right) \nonumber \\
\Rightarrow \; T & = &  \underbrace{\exp \left(\Delta_L \, \frac{\delta}{\delta\Delta_S} \right)}_{T_{FSI}} \;\; T_S \;\;\underbrace{\exp \left(-\, \Delta_L \, \frac{\delta}{\delta\Delta_S} \right)}_{T_{ISI}} \, .
\end{eqnarray}
It is important to realise that the matrix $\frac{\delta}{\delta\Delta_S}$ has to act like a functional derivative\footnote{One easily can proof, that there exists {\em for a finite dimensional matrix} $\Delta_S$ no matrix ${\cal M}$ such, that the commutator $\left[ {\cal M} \, ,\, \Delta_S \right]$ yields the unit matrix.}. On the other hand the unitarity condition $T^{-1}_{FSI} = T^+_{FSI}$ enforces $\left[ \frac{\delta}{\delta\Delta_S}\; ,\; \Delta_L \right] = 0$. 
We can learn from the ``naive'' considerations presented above 3 points\footnote{It is left to future work to investigate the constraints of analyticity properties of 5-particle amplitudes \cite{alv73} like the amplitude of the process $NN\rightarrow NNX$ to enhancement factors. For the discussion of 2- and 3-body unitarity in elastic T- and S-matrices I refer the reader e.g.\ to \cite{wal74,gar01}.}: \\

\noindent {$\bullet$} It is a very difficult --- most probably only approximately soluble --- task to systematically construct enhancement factors, which fulfil simultaneously the requirements $\left[ \frac{\delta}{\delta\Delta_S}\, ,\, \Delta_S \right] = 1$ and $\left[ \frac{\delta}{\delta\Delta_S}\, ,\, \Delta_L \right] = 0$. The first commutator can be fulfilled only on a functional or operator level.

\noindent {$\bullet$} $T_{FSI}$ and $T_{ISI}$ are related ($T_{FSI} = T^{-1}_{ISI}$) and constrained by unitarity (see also \cite{kle00b,nis98}).

\noindent {$\bullet$} The construction of enhancement factors is strongly based on $\Delta_S$, i.e.\ {\em enhancement factors do depend on short-range interactions}\footnote{With respect to the last point, I have to admit, that V.\ Baru et al.\ \cite{bar00,bar01} are right, if they state \cite{bar01}:
``The absolute magnitude of FSI effects depends on the momentum transfer (or on the mass of the produced meson) and hence is not universal.'' They deduce from a simple rescattering graph in $NN\rightarrow NNX$, that the enhancement factor only for heavy produced mesons seem to be universally factorisable. Yet if they state, that ``$\ldots$ FSI effects cannot be factorised from the production amplitude $\ldots$'', they ignore the possibility, that for every short-ranged sub-amplitude in the process $NN\rightarrow NNX$ the enhancement factor may be constructed individually in a matrix form discussed. Of course it seems more attractive to treat the process $NN\rightarrow NNX$ in a Faddeev formalism rather than factorising short- and long-ranged interactions. Yet for processes involving more particles (especially with spin) it seems at present beyond the numerical capabilites to avoid a Watson-Migdal formalism.}.
\section{Energy expansion and behaviour of the total cross section at close to threshold}
\subsection*{General idea}
The total cross section of $NN\rightarrow NN\phi$ is given by  (defining $P:= p_1 + p_2$, $P^\prime := p_{1^\prime} + p_{2^\prime} + k_\phi$ and the flux factor ${\cal F} (s) := 2 \, \sqrt{\lambda (s; m^2_1, m^2_2)}$):
\begin{eqnarray} \sigma (s) & = &  \frac{{\cal S}}{{\cal F} (s)} \; 
\int\!\frac{d^{\,3}p_{1^\prime}}{(2\pi )^3\, 2\,\omega_N
(|\vec{p}_{1^\prime}|)}
\frac{d^{\,3}p_{2^\prime}}{(2\pi )^3\, 2\,\omega_N
(|\vec{p}_{2^\prime}|)}
\frac{d^{\,3}k}{(2\pi )^3\, 2\,\omega_\phi (|\vec{k}_\phi |)} \nonumber \\
 & \cdot & (2\pi )^4 \; \delta^4 (P^\prime - P) \;\; \overline{|T_{fi}|^2} \; .
\end{eqnarray}
The symmetry factor ${\cal S}$ is $\frac{1}{2!}$ for two identical nucleons in the final state, otherwise 1. $\overline{|T_{fi}|^2}$ denotes e.g.\ spin averaging, i.e. $\frac{1}{4} \; 
 \sum_{s_1,s_2,s_{1^\prime},s_{2^\prime}} \; |T_{fi}|^2$. As for this reaction $\sigma(s)$ at threshold vanishes exactly\footnote{This is not true for nucleon nucleus collisions, which allow also subthreshold production cross section due to e.g.\ the Fermi motion of the nucleons in the nucleus.} it is suitable to perform an energy expansion of the $\sigma(s)$ in a sense of a Taylor expansion in terms of the small energy dependent parameters $\eta_\phi$ or $Q_{cm}$. The argument of the phasespace integration $|T_{fi}|^2 = |T_{FSI} \, T^{(0)} \, T_{ISI}|^2$ can be expressed as a function of the 5 independent Lorentz-invariants $s$, $s_1$, $s_2$, $t_1$, $t_2$. As discussed above, close to threshold $T_{FSI}$  will mainly depend on $\kappa_{12}$, $\kappa_{23}$, $\kappa_{31}$, i.e. $T_{FSI}\simeq T_{FSI}(s,s_1,s_2)$. $T_{ISI}$ depends mainly on $\kappa_0$, i.e. $T_{ISI} \simeq T_{ISI} (s)$. If the propagators in the short-ranged T-matrix $T^{(0)}(s, s_1, s_2, t_1, t_2)$ are not too close to their pole and particles are mainly produced through nucleons or S-wave resonances, it is a good approximation to treat close to threshold the T-matrix as a constant\footnote{This approximation yielding very compact analytical results has been shown to work quite well for the reactions $pN\rightarrow pNX$ with $X=\eta,\pi^0$ (for details see e.g.\ \cite{kle99a,kle98}). A corresponding approximation has been later applied to similar reactions by \cite{ber99}.}, i.e.  $T^{(0)} \simeq T^{(0)}(s^{\,thr}, s^{\,thr}_1, s^{\,thr}_2, t^{\,thr}_1, t^{\,thr}_2)$ with 
\begin{eqnarray} & & s^{\,thr} = (m_{1^\prime} + m_{2^\prime} + m_\phi)^2 , \; 
s_1^{\,thr} = (m_{1^\prime} + m_{2^\prime})^2 , \; s_2^{\,thr} = (m_{2^\prime} + m_\phi)^2 , \nonumber \\
 & & t_1^{\,thr} = m^2_1 - \, m_{1^\prime} \, (m_{2^\prime} + m_\phi) , \; 
 t_2^{\,thr} = m^2_2 - \, m_\phi \,(m_{1^\prime} + m_{2^\prime}) \; .
\end{eqnarray}
To improve this approximation one can perform a Taylor expansion of $T^{(0)}(s, s_1, s_2, t_1, t_2)$ in the variables $\delta s := s - s^{\,thr}$, $\delta s_1 := s_1 - s^{\,thr}_1$, $\delta s_2 := s_2 - s^{\,thr}_2$, $\delta t_1 := t_1 - t^{\,thr}_1$, $\delta t_2 := t_2 - t^{\,thr}_2$, or simply a Taylor expansion in the variable $\delta s := s - s^{\,thr}$, i.e.\footnote{This kind of a energy expansion ot $T^{(0)}$ is (in terms of the variable $\eta_\phi$) for the first time taken into account by \cite{fae01}. The problem is, that the Taylor coefficients are derived from onshell meson nucleon scattering data and not from any theoretical meson nucleon potential.}:
\begin{eqnarray}
\lefteqn{T^{(0)}(s, s_1, s_2, t_1, t_2) \; = } \nonumber  \\
 & = & \sum\limits_{n=0}^\infty \frac{(\delta s)^n}{n!}\frac{\partial^n T^{(0)}(s^{\,thr}, s_1, s_2, t_1, t_2)}{\partial s^{\,thr\; n}} =: \sum\limits_{n=0}^\infty (\delta s)^n  \; T_n^{(0)} (s_1, s_2, t_1, t_2)  . \,\quad\;
\end{eqnarray}
Instead of performing a Taylor expansion in $\delta S$ or $Q_{cm}$ it is more suitable to perform a Taylor expansion in the dimensionless variable $\eta_\phi$, which is related to $s$ and $Q_{cm}$ by ($2\,\mu :=(m_{1^\prime}+m_{2^\prime})/m_\phi $):
\begin{equation} \sqrt{\frac{s}{m^2_\phi}} \; = \; \frac{Q_{cm}}{m_\phi} + 1+ 2\,\mu \; = \; \sqrt{\eta_\phi^2 + 1} \; + \; 
\sqrt{\eta_\phi^2 + (2\,\mu)^2} \; . \end{equation}
Pulling all factors depending only on $s$ in front of the phasespace integration and selecting one specific spin-isospin channel one obtains for $\sigma(s)$:
\begin{eqnarray} \sigma (s) & \simeq & 
 \frac{1}{(2\pi)^5} \; \frac{{\cal S}}{{\cal F} (s)} \; |T_{ISI} (s)|^2 \;\;\sum\limits^\infty_{n=0} \;(\delta s)^n
 \, \cdot
\nonumber \\
 & \cdot & 
\int\!\frac{d^{\,3}p_{1^\prime}}{2\,\omega_N
(|\vec{p}_{1^\prime}|)}
\frac{d^{\,3}p_{2^\prime}}{2\,\omega_N
(|\vec{p}_{2^\prime}|)}
\frac{d^{\,3}k}{2\,\omega_\phi (|\vec{k}_\phi |)} \; \delta^4 (P^\prime - P) \;\cdot \nonumber \\
 & & \nonumber \\
 & \cdot & |T_{FSI} (s,s_1,s_2) \; T_n^{(0)} (s_1,s_2,t_1,t_2)|^2 \; . 
\end{eqnarray}
The energy behaviour of $\sigma (s)$ is determined on one hand by the $s$-dependent function in front of the phasespace integral consisting of the flux factor ${\cal F}$, the initial state enhancement factor $T_{ISI}$ and the factor $(\delta s)^n$, on the other hand by the energy dependence of the phasespace integral itself, which is affected strongly by $T_{FSI}$. In the following I perform the threshold energy expansion of the various factors in terms of the varibable $\eta_\phi$ separately. The flux factor behaves very close to threshold like a constant, while the dominant term in the power expansion of $\delta s$ is the term with $(\delta s)^0$, as all terms with $n>0$ will be of higher order in $\eta_\phi$.
\subsection*{Treatment of ISI}
The energy expansion of $T_{ISI} (s) = 1 + i \,  ( e^{i\,\delta (\kappa_0)} \, \sin \delta (\kappa_0)) (1-\,P(\kappa_0) ) $ is discussed in \cite{kle99c}. I will give here merely the general ideas. Even if many people think, the offshell effects in the initial state are very small, i.e. $P(\kappa_0) \simeq 0$, the relation $T_{FSI} = T^{-1}_{ISI}$ shows, that this is most probably not true. Therefore I perform the Taylor expansion $P(\kappa_0) = P^{(0)} + P^{(2)} \, \eta^2_\phi + O(\eta^4_\phi)$. As phaseshifts\footnote{I treat here phaseshifts as complex numbers containing also eventual inelasticities!} are given in terms of the kinetic Lab-energy $T_{Lab} = (s - (m_1 + m_2)^2)/(2\,m_2)$ (particle 2 at rest!), one performs the following Taylor expansion of the data around threshold:
\begin{eqnarray} \delta (T_{Lab}) & = & \delta (T^{\,thr}_{Lab}) + \delta^{\;\prime} (T^{thr}_{Lab}) \; (T_{Lab} - T^{\,thr}_{Lab}) \; + \; O ((T_{Lab} - T^{\,thr}_{Lab})^2)   \nonumber \\
 & = & \delta (T^{\,thr}_{Lab}) + \delta^{\;\prime} (T^{thr}_{Lab}) \; \frac{m_\phi^2 \, \mu}{m_2} \; \left( 1 + \frac{1}{2\,\mu} \right)^2 \; \eta^2_\phi + \; O(\eta^4_\phi) \nonumber \\
 & = & \delta^{\;(0)} + \delta^{\;(2)} \; \eta^2_\phi + \; O(\eta^4_\phi) \, . 
\end{eqnarray}
($T^{\,thr}_{Lab} = ((m_{1^\prime}+m_{2^\prime}+m_\phi)^2-\,(m_1+m_2)^2) / (2 \, m_2)$). Combining all we get:
\begin{eqnarray} \lefteqn{T_{ISI} (s) \; = \; 1 \,+ i \;  ( e^{i\,\delta^{\,(0)}} \, \sin \delta^{\,(0)}) (1-\,P^{\,(0)}) \; +} \nonumber \\
 & + & i\; \left( e^{2\, i\,\delta^{\,(0)}} \, \delta^{\,(2)} \,(1-\,P^{\,(0)}) -  \,P^{\,(2)} \, e^{i\,\delta^{\,(0)}} \, \sin \delta^{\,(0)} \right) \, \eta^2_\phi + \; O(\eta^4_\phi) \, . \quad  
\end{eqnarray}
The $NN$ phaseshifts at the threshold of $pp\rightarrow pp\,X$ are badly known for mesons heavier than the pion. The accuracy of the determination of high energy phaseshifts limits a quantitative theoretical calculation of $\sigma(s)$. ISI --- close to threshold --- don't seem to affect the energy dependence of the cross section. Their effect is just a multiplication of the cross section by some number\footnote{Assuming $P(\kappa_0)\simeq 0$ and taking $\delta^{\;(0)}\approx - 60^o$ and $\delta^{\;(2)}\approx 0$ from the VPI-phaseshift analysis \cite{vpi1} one obtaines for $pp\rightarrow pp\eta$ close to threshold $|T_{ISI} (s^{thr})|\approx 1/2$, i.e. a reduction of $\sigma (s)$ by a factor of about 1/4.} (see also e.g.\ \cite{bat97}).
\subsection*{Treatment of FSI}
From considerations presented above I conclude, that in leading order the energy-, i.e.\ $\eta_\phi$-dependence of $\sigma(s)$ close to threshold is determined by the energy-, i.e.\ the $\eta_\phi$-dependence of the phasespace integral over a FSI-dependent integrand. I.e.\ the problem is now to determine the threshold energy- or $\eta_\phi$-dependence of the following FSI-modified phasespace integral:
\begin{eqnarray} R^{\,FSI}_3 (s) & := & 
\int\!\frac{d^{\,3}p_{1^\prime}}{2\,\omega_{1^\prime}
(|\vec{p}_{1^\prime}|)}
\frac{d^{\,3}p_{2^\prime}}{2\,\omega_{2^\prime}
(|\vec{p}_{2^\prime}|)} 
\frac{d^{\,3}k}{2\,\omega_\phi (|\vec{k}_\phi |)} \;\; \delta^4 (P^\prime - P) \; \cdot \nonumber \\
 & \cdot & |T_{FSI} (s,s_1,s_2) \; T_n^{(0)} (s_1,s_2,t_1,t_2)|^2 \; . 
\end{eqnarray}
The phasespace integration can be rewritten into a restricted integration over four Lorentz-invariants $s_1$, $s_2$, $t_1$, $t_2$, or into a restricted integration over three of these invariants (e.g. $s_1$, $s_2$, $t_1$) and one unrestricted angle, while the fourth of these invariants (e.g.\ $t_2$) is expressed in terms of the three and the angle (see e.g. \cite{byc73}). Remembering, that $|T_n^{(0)} (s_1,s_2,t_1,t_2)|$ close to threshold is nearly constant, I want to investigate now only the threshold energy dependence of the phasespace integral with a integrand $f$ being a function of $s_1$, $s_2$ and $s$ only\footnote{The investigation of the general case is devoted to future research.}, i.e.\  ($S:=s/m^2_\phi$, $S_1:=s_1/m^2_\phi$, $S_2:=s_2/m^2_\phi$):
\begin{equation} R^{\,FSI}_3 (s) \; \simeq \; \int\!\frac{d^{\,3}p_{1^\prime}}{2\,\omega_{1^\prime}
(|\vec{p}_{1^\prime}|)}
\frac{d^{\,3}p_{2^\prime}}{2\,\omega_{2^\prime}
(|\vec{p}_{2^\prime}|)}
\frac{d^{\,3}k}{2\,\omega_\phi (|\vec{k}_\phi |)} \;
 \, \delta^4 (P^\prime - P) \;\, f \left( S,S_1,S_2\right) \, . \label{gen1}
\end{equation}
The case considered most in $NN\rightarrow NNX$ close to threshold is the scenario, when $T_{FSI}$ is described as a sum of T-matrices describing the FSI of the 2-particle subsystems $1^\prime 2^\prime$, $2^\prime 3^\prime$ and $1^\prime 3^\prime$ in the final state only, i.e.\ $T_{FSI} (1^\prime 2^\prime 3^\prime) \simeq T_{FSI} (1^\prime 2^\prime) + T_{FSI} (2^\prime 3^\prime) + T_{FSI} (3^\prime 1^\prime) - 2$. This scenario in the OWMA is equivalent of using only the lowest order terms in the Faddeev expansion of the 3-particle elastic T-matrix in the final state. In an incoherent language\footnote{The coherent case for $m_{1^\prime} \not= m_{2^\prime}$ has still to be investigated.} with the correct limit for no FSI this would yield $|T_{FSI} (1^\prime 2^\prime 3^\prime)|^2 \simeq |T_{FSI} (1^\prime 2^\prime)|^2 + |T_{FSI} (2^\prime 3^\prime)|^2 + |T_{FSI} (3^\prime 1^\prime)|^2 - 2$. Hence in this case we have $f \left( S,S_1,S_2\right) \simeq f_{12} (\kappa_{12}) + f_{23} (\kappa_{23}) + f_{31} (\kappa_{31})$, i.e.:  
\begin{eqnarray} R^{\,FSI}_3 (s) & \simeq & \int\!\frac{d^{\,3}p_{1^\prime}}{2\,\omega_{1^\prime}
(|\vec{p}_{1^\prime}|)}
\frac{d^{\,3}p_{2^\prime}}{2\,\omega_{2^\prime}
(|\vec{p}_{2^\prime}|)}
\frac{d^{\,3}k}{2\,\omega_\phi (|\vec{k}_\phi |)} \;
 \, \delta^4 (P^\prime - P) \; \cdot \nonumber \\
 & & \cdot \; (f_{12} (\kappa_{12}) + f_{23} (\kappa_{23}) + f_{31} (\kappa_{31})) \, .
\end{eqnarray}
If the functions $f_{ij}(\kappa_{ij})$ can be Taylor expanded, i.e. $f_{ij}(\kappa_{ij})=\sum\limits_\alpha \,c^{\,(\alpha)}_{ij} \, \kappa^{\,\alpha}_{ij}$, then the threshold behaviour of $R^{FSI}_3(s)$ is given by \cite{kle00b}: 
\begin{eqnarray} \lefteqn{R^{\,FSI}_3 (s)\; \simeq \; \sum\limits_\alpha\, \frac{\pi^2 \, m^{\,\alpha + 2}_\phi}{2^{\,\alpha +1} \, (2 \mu )} \, I_\alpha \; \cdot} \nonumber \\
 & \cdot &  \left( c^{\,(\alpha)}_{12} +  c^{\,(\alpha)}_{23}  \left(\frac{m_\phi\, J_+}{m_{1^\prime}}\right)^{\,\alpha + 4} + c^{\,(\alpha)}_{31} \left(\frac{m_\phi\, J_-}{m_{2^\prime}}\right)^{\,\alpha + 4} \right) \left( \eta_\phi^{\,\alpha + 4} + O(\eta_\phi^{\,\alpha + 6}) \right) \nonumber \\
\end{eqnarray}
using the definitions $\Delta := \sqrt{1 - (m_{1^\prime}-m_{2^\prime})^2 (m_{1^\prime}+m_{2^\prime})^{-2}}\;$ and $I_\alpha := \int^{\,1}_{0}\!du \; \sqrt{u \; ( (1 + 2 \mu ) \Delta^2 (1-u) )^{\alpha + 1}}$ and $J_\pm:= \frac{1}{2} \; \sqrt{2\pm 2 \, \sqrt{1-\Delta^2} + (2\mu) \Delta^2\,}$. Considering these results it is easy to see\footnote{Using $\int^{\,1}_{0}\!du \; \sqrt{u \, (1-u)} = \pi/ 8$ one easily can deduce the expression for the free phasespace, i.e.\ $R_3(s) = \frac{\pi^3}{16} \; m_\phi^2\;  \Delta \; \frac{\sqrt{1 + 2\, \mu}}{2\,\mu} \; \eta^4_\phi + O(\eta^6_\phi)\, .$}, that the free 3-body phasespace (i.e.\ $f(S,S_1,S_2)=1$) is proportional to $\eta^4_\phi$.
Finally I want to discuss the more general case (\ref{gen1}), yet for equal masses $m_{1^\prime}=m_{2^\prime}=m_N$, i.e. $\Delta = 1$. In this case $R^{FSI}_3(s)$ can be rewritten as \cite{kle99a}:
\beqa \lefteqn{R^{FSI}_3 (s) \; = } \nonumber \\
 & \stackrel{!}{=} &
\frac{1}{2\, \sqrt{S}} \,\, \pi^2 \, m_\phi^2 \; \eta^2_\phi \int_{\; 0}^{\;1}
 du \;\; \frac{B}{\sqrt{\eta^2_\phi\, u+1}}  
 \int_{\; 0}^{\;1} dv \;\; \frac{1}{2} \;   
 [ \; 
 f(\, S \, , \, S_1 \, , \, A + B \, v \, )  \nonumber \\
 & & \;\;\makebox[6.9cm]{} + \,
 f(\, S \, , \, S_1 \, , \, A - B \, v \, ) 
 \; ] \qquad
\eeqa
with $\sqrt{S} \; \stackrel{!}{=} \; \sqrt{\eta^2_\phi + 1} \; + \; 
 \sqrt{\eta^2_\phi + (2\,\mu )^2}$, $S_1 \; \stackrel{!}{=} \; S + 1 - \, \sqrt{4\, S} \;  \sqrt{\eta^2_\phi \, u + 1}$, $A \; := \; \frac{1}{2} \; [S - S_1 + 2\,\mu^2 +1]$ and
 $B \; := \;  \sqrt{S \, \eta^2_\phi \, u \;\, (S_1 - (2\,\mu )^2)/S_1}$. Remembering that $S$, $S_1$, $A$ and $B$ are functions of $\eta^2_\phi$ and $u$, i.e. $S=S(\eta^2_\phi )$, $S_1=S_1(u,\eta^2_\phi )$, $A=A(u,\eta^2_\phi )$ and
$B=B(u,\eta^2_\phi )$, it is easy to see, that the integrand can be Taylor expanded in the variable $\eta^2_\phi$. Afterwards one has to calculate analytically/numerically the  $u$-/$v$-integrals in the Taylor coefficients.
Using {\sl Mathematica${}_{{}_{\mbox{\footnotesize $\bigcirc$}}\!\!\!\!\!\!\!\!\:\: 
{}_{{}_{\mbox{\tiny R}}}}$} it is no problem handle Taylor expansion and integrations.
\subsection*{The Coulomb problem and how to surround it}
Upto now we learned, that we easily can deduce the threshold $\eta_\phi$-dependence of the total cross section by performing a Taylor expansion of $|T_{FSI}\; T^{(0)}|^2$ in the relative wavenumbers, say e.g.\ $\kappa$. This is always possible for well behaved long-ranged potentials, which yield e.g. for the S-wave the standard ERE of the following form (see e.g.\ p.\ 466 in \cite{noy72}, \cite{squ1} or p.\ 12ff in \cite{bro2}):
\begin{equation} \kappa \, \cot \delta_0(\kappa) = -\,\frac{1}{a} + \frac{1}{2} \; r \; \kappa^2 - \, P \; r^3 \, \kappa^4 + Q\;r^5 \, \kappa^6 + \ldots \end{equation}
If we keep only the leading two terms in the ERE, i.e. only $\kappa \, \cot \delta_0(\kappa) = -\,a^{-1} + r \; \kappa^2/2$ we call this ERE ``shape independent''. Specific properties of the long- and/or short-ranged potential under consideration can absorbed in higher order ``shape dependent'' terms of the expansion. In 1959 M.\ Cini, S.\ Fubini, A.\ Stanghellini (CFS) \cite{cin1} included the singularity  of an One-Pion Exchange (OPE) potential at $\kappa^2 = - m^2_\pi/4$ to the ERE by (see also \cite{noy72}):
\begin{equation} \kappa \cot \delta_0 (\kappa) \; = \; - \; \frac{1}{a} \; + \; \frac{1}{2} \; r \; \kappa^2  -\, \frac{p\;\kappa^{\,4}}{1 + q\; \kappa^2}\,.
\end{equation}
The absorption of more pathologic singularities like the essential singularity of the one-photon exchange (at $\kappa^2=0$) in the Coulomb interaction is performed within a formalism using a ``Modified Effective Range Function'' (MERF) $K^M_L(\kappa^2) := A_L(\kappa^2) + B_L(\kappa^2) \; \kappa^{2\,L + 1} \cot \delta_L (\kappa)$ \cite{hae2,hae1}\footnote{As shown in \cite{hae2} the functions $A_L(\kappa^2)$ and $B_L(\kappa^2)$ are related to the Jost functions and Jost solutions of the potential problem under consideration.}. The corresponding ``Modified Effective Range Expansion (MERE)'' is given by:
\begin{equation} B_L(\kappa^2) \; \kappa^{2\,L + 1} \, \cot \delta_L(\kappa) + A_L(\kappa^2) = -\,\frac{1}{a} + \frac{1}{2} \; r \; \kappa^2 + \ldots \end{equation}
It is easy to see, how the MERE enters $T_{FSI}$ described e.g.\ in (\ref{eqx3}), i.e.:
\begin{eqnarray} T_{FSI} & = & \frac{\kappa^{2\,L + 1} (\cot \delta_L (\kappa) - i \, P (\kappa))}{\kappa^{2\,L + 1} (\cot \delta_L (\kappa) - i )\qquad\;} \nonumber \\
 & \stackrel{!}{=} & \frac{-\,\frac{1}{a} + \frac{r}{2} \, \kappa^2 - \ldots - A_L (\kappa) - i \,\kappa^{2\,L + 1} \, B_L (\kappa) \, P (\kappa)\;\;}{-\,\frac{1}{a} + \frac{r}{2} \, \kappa^2 - \ldots - A_L (\kappa) - i \, \kappa^{2\,L + 1} \, B_L (\kappa)\qquad\;\;\;} \, .
\end{eqnarray}
So, if $|T_{FSI}|^2$ can be Taylor expanded, we immediately can derive the threshold energy dependence of the total cross section. To handle e.g.\ the $pp$-FSI in $pp\rightarrow pp\pi^0$, we have to use a Coulomb Modified Effective Range Expansion (CMERE) suggested by G.\ Breit et al.\ (1936) \cite{bre1} and extensively investigated e.g.\ by \cite{won1,noy3,aus82,hae1} \footnote{For the theory of $pp$-scattering see e.g. also \cite{lan1,lan3,jac1,bre3}, p.\ 49ff in \cite{bro2},\cite{che1,rei68}.
For the quantum-mechanical Coulomb problem and the ERE see e.g. \cite{yos1,tay1,tay2}, p.\ 133ff in \cite{joa83},\cite{hof1,bet3}, p.\ 323 in \cite{ebe1}, p.\ 285 in \cite{dum1}, p.\ 54ff in \cite{bro2}, p.\ 161ff in \cite{omn1}, p.\ 16 in \cite{lan2}.)}. For S-wave scattering the CMERE yields $B_0(\kappa) = {\cal C}_0^{\,2} (\eta ) := 2\,\pi\,\eta /(\exp (2\,\pi\,\eta) -1)$ (``Coulomb penetration factor'', ``Gamov factor'') and $A_0(\kappa) = 2 \, \kappa \, \eta \, H(\eta)$ with the ``Sommerfeld parameter'' $\eta := \alpha /v_{pp} = \alpha \, \sqrt{(\kappa^2+(m_{1^\prime}m_{2^\prime}/(m_{1^\prime}+m_{2^\prime}))^2)/\kappa^2}$ and $\alpha \simeq 1/137$ \footnote{Note, that $H(\eta) = - \gamma - \ln \eta + \sum^\infty_{s=1} \eta^2 /(s\,(s^2 + \eta^2))$ and $2 \, \kappa \, \eta \simeq \alpha \,m_p = (28.82\,\mbox{fm})^{-1}$!}. In the 2-potential formalism the total phaseshift $\delta_0(\kappa)$ can be decomposed into a sum of the Coulomb phaseshift  $\sigma_0 (\kappa)$ ($\exp(2\,i\,\sigma_0) = \Gamma (1+i\,\eta)/\Gamma (1-i\,\eta)$) and the phaseshift $\delta^c (\kappa)$ of the short-ranged potential in the presence of the Coulomb potential, i.e. $\delta_0 (\kappa) = \sigma_0 (\kappa) + \delta_0^{\,c} (\kappa)$. In the CFS-CMERE derived by D.Y.\ Wong et al.\  (1964) \cite{won1}, which is based on a 2-potential formalism, one obtains (see e.g.\ also p.\ 468 in \cite{noy72}, p.\ 1869 in \cite{won1}, or \cite{jac1,ber9,moa96,shy99,rei68}):
\begin{equation} {\cal C}_0^{\,2} (\eta ) \, \kappa \cot \delta_0^{\,c} (\kappa) + 2 \, \kappa \, \eta \, H(\eta) \; = \; - \; \frac{1}{a^{\,c}} \; + \; \frac{1}{2} \; r^{\,c} \; \kappa^2 - \frac{p^{\,c} \;\kappa^{\,4}}{1 + q^{\,c} \; \kappa^2} \, .\label{eqcfs}
\end{equation}
The construction of an enhancement factor described e.g.\ in (\ref{eqx3}) is straightforward:
\begin{eqnarray} \lefteqn{T_{FSI} \quad = \quad 1 + \frac{1}{2} \left( e^{\,2\,i\,\delta_0 (\kappa)} - 1 \right) \Big( 1- P(\kappa )\Big) \quad =} \nonumber \\
 & = &  1 + \frac{1}{2} \left( e^{\,2\,i\,\sigma_0 (\kappa)} \, e^{\,2\,i\,\delta^{\,c}_0 (\kappa)} - 1 \right) \Big( 1- P(\kappa )\Big)  
 \nonumber \\
 & = &  1 + \frac{1}{2} \left( e^{\,2\,i\,\sigma_0 (\kappa)} \, \frac{\kappa \, (\cot\,\delta^{\,c}_0 (\kappa)+i)}{\kappa \, (\cot\,\delta^{\,c}_0 (\kappa)-i)} - 1 \right) \Big( 1- P(\kappa )\Big)  
 \nonumber \\
 & = & 1 + \frac{1}{2} \left( e^{\,2\,i\,\sigma_0 (\kappa)} \, \frac{- \; \frac{1}{a^c} \; + \ldots  - 2 \, \kappa \, \eta \, H(\eta) + \, i \,\kappa \; {\cal C}_0^{\,2} (\eta )}{- \; \frac{1}{a^c} \; +  \ldots  - 2 \, \kappa \, \eta \, H(\eta) - \, i \,\kappa \; {\cal C}_0^{\,2} (\eta )} - 1 \right) \Big( 1- P(\kappa )\Big) . \nonumber \\
\label{eqx5} \end{eqnarray}
Due to essential singularities appearing in the differentiation of $\sigma_0 (\kappa)$ and ${\cal C}_0^{\,2} (\eta )$ in the limit $\kappa\rightarrow 0$ it is for a Coulomb potential not possible to perform a Taylor expansion of $|T_{FSI}|^2$ in the variable $\kappa$, in order to determine the energy behaviour of the total cross section close to threshold. Yet the problem can be surrounded in the following way: first one regularises the Coulomb potential according to ($\mu ,\varepsilon > 0$)($g := \alpha\, m_p = \lim\limits_{\kappa\rightarrow 0} 2 \, \kappa \, \eta$):
\begin{eqnarray} U(r) = g\; \frac{e^{-\,\mu \, r}}{r+\varepsilon}
 & = & g \int^\infty_0 dt \; \theta (t-\mu) \; e^{\,(\mu - t)\,\varepsilon}  \; e^{-\,t \, r} \nonumber \\
 & \stackrel{p.i.}{=} & g \int^\infty_0 dt \; \left[ \, \delta(t-\mu) - \varepsilon \; \theta (t-\mu) \; \right] \; e^{\,(\mu - t)\,\varepsilon} \;\,  \frac{e^{-\,t \, r}}{r} \, . \; \quad 
\end{eqnarray}
As the potential fulfils $|\int^\infty_0 dr \, r \, U(r)|<\infty$ and is of Yukawa type, one can determine by the method described in \cite{trl80} the corresponding Jost solution\footnote{The Jost solution $f_0(\kappa,r)$ is obtained to be (Ei$ (- z) := - \int^\infty_z \exp(- t)/t$ for $z>0$):
\begin{eqnarray} \lefteqn{f_0(\kappa,r) \; = } \nonumber \\
 & = & e^{\,i\,\kappa\,r} \Big\{ 1 + \sum\limits_{n=1}^\infty \left( g\,\exp(\mu\,\varepsilon) \right)^n
 \; \int^\infty_{n\mu} d\tau_n \; \frac{\exp (- \,\tau_n (r+\varepsilon))}{\kappa^2 + (-i \,\kappa + \tau_n)^2} \nonumber \\
 & & \qquad \qquad \qquad \cdot \; \prod\limits^{n-1}_{\nu = 1} \left[ \; \int^\infty_{\nu\mu} d\tau_\nu \; \frac{1}{\kappa^2 + (-i \,\kappa + \tau_\nu)^2} \; \right] 
\Big\} \nonumber \\
& = & e^{\,i\,\kappa\,r} \Big\{ 1 + \sum\limits_{n=1}^\infty \left( \frac{g\,\exp(\mu\,\varepsilon)}{2\,\kappa\,i} \right)^{\!n} \!\!
 [ \, \mbox{Ei} (- (r+\varepsilon) \,n\mu) - e^{-\,2\,i\,\kappa (r+\varepsilon)} \, \mbox{Ei} (- (r+\varepsilon) (n \mu - 2\,i\,\kappa ))] \nonumber \\
 & & \qquad \qquad \qquad \cdot \; \prod\limits^{n-1}_{\nu = 1} \left[ \frac{1}{2} \, \ln \left(\frac{(\nu \mu)^2}{(\nu \mu)^2 + 4\,\kappa^2} \right)  + i \, \left( \frac{\pi}{2} - \arctan \frac{\nu\mu}{2\,\kappa} \right) \right] 
\Big\} \, . 
\end{eqnarray}.} $f_0(\kappa,r)$ and Jost function $f_0(\kappa):=f_0(\kappa,0)$. As the potential is analytic at $r=0$, it is then possible --- assuming $\kappa$ to be real positive --- to apply the method in section IV of \cite{hae1} to determine from $f_0(\kappa,r)$ and $f_0(\kappa)$ a MERF $K^M_0(\kappa^2)$ for the regularised potential by ($f^{\prime}_0(\kappa,0) := \lim\limits_{r\rightarrow 0} \partial f_0(\kappa,r)/ \partial\, r$):
\begin{equation} K^M_0(\kappa^2) = \frac{f^{\,\prime}_0 (\kappa,0)}{f_0 (\kappa,0)} + \frac{1}{|f_0 (\kappa)|^2} \;\,\kappa \; \Big(\cot \delta^M_0 (\kappa) - i \Big) = - \, \frac{1}{a} + \frac{1}{2} \; r \; \kappa^2 + \ldots 
\end{equation}
The regularised enhancement factor is now obviously obtained from (\ref{eqx5}) by the replacements $e^{\,2\,i\,\sigma_0 (\kappa)} \rightarrow f^\ast_0(\kappa^\ast)/f_0(\kappa)$, $ {\cal C}_0^{\,2} (\eta ) \rightarrow 1/|f_0(\kappa)|^2$, $2 \, \kappa \, \eta \, H(\eta) \rightarrow f^{\,\prime}_0 (\kappa,0) /f_0(\kappa,0) - i \, \kappa / |f_0(\kappa)|^2$, $\delta_0^{\,c}(\kappa)\rightarrow \delta^M_0 (\kappa)$. The result is:
\begin{eqnarray} T_{FSI} 
 & = & 1 + \frac{1}{2} \left( \frac{f^\ast_0(\kappa)}{f_0(\kappa)} \, \frac{\left(- \; \frac{1}{a} \; + \ldots  - \frac{f^{\,\prime}_0 (\kappa,0)}{f_0(\kappa,0)} + \frac{2\,i\,\kappa}{|f_0(\kappa)|^2}\right)}{\left(- \; \frac{1}{a} \; + \ldots  - \frac{f^{\,\prime}_0 (\kappa,0)}{f_0(\kappa,0)}\right)\qquad\qquad \;} - 1 \right) \Big( 1- P(\kappa )\Big) . \nonumber \\
\label{eqx7} \end{eqnarray}
Remember, that $f^\ast_0(\kappa)/f_0(\kappa) = |f_0(\kappa)|^2/(f_0(\kappa))^2$. Now $|T_{FSI}|^2$ can be Taylor expanded and the threshold energy dependendence of the total cross section be determined. The Coulomb FSI is obtained by $\mu\rightarrow +0$ and $\varepsilon\rightarrow +0$.
Eq.\ (\ref{eqx7}) can be also used, to construct enhancement factors from Jost functions obtained for singular potentials by the WKB approach \cite{sof99}.
\subsection*{Effective Range Expansion and astrophysics}
One valuable source \cite{bay00,sat99} for sophisticated Coulomb destorted wavefunctions at low energies and CMEREs to be used in $pp$ scattering is astrophysics, which is due to a interesting correspondence between $NN\rightarrow NNX$ close to threshold and theory of radiative capture processes in astrophysics\footnote{Compare e.g. the reaction $pp\rightarrow d\,\pi^+$ close to threshold and the capture process $p A\rightarrow (A+1) \gamma$. In the first process the final state $d\,\pi^+$ is nearly at rest, while in the second process the initial state $pA$ is nearly at rest. The capture process behaves like time reversed compared to a threshold production process. Both processes have enhancement factors. In $NN\rightarrow NNX$ we have $T_{FSI}$, in the capture process we have the astrophysical S-factor $S(E)$. The T-matrix elements look very similar: for $NN\rightarrow NN\pi$ we have $T_{fi} \sim \int dr \,u_f \, j_1 (|\vec{k}_\pi|\, r) u_i$ with $u_f$ at $E\simeq 0$, while in the capture process we have $T_{fi} \sim \int dr \, u_f \, r^\lambda \, u_i$ with $u_i$ at $E\simeq 0$. The only difference is, that the excitation energy of the meson production process is high, while in the capture process all particles stay nearly onshell.}. 
\section{$|T_{FSI}|$ onshell beyond a threshold expansion} \label{sec5}
\subsection*{The FSI model of F\"aldt and Wilkin}
A particularly interesting existing nonrelativistic theoretical model for the description of FSI is the model developed by F\"aldt and Wilkin \cite{fal96a,fal96b,fal96c,fal97}. It does not only predict the leading terms in the $\eta_\phi$-Taylor expansion of the total cross section close to threshold, it gives expressions describing the threshold energy behaviour of the total cross section {\em to all orders} in $\eta_\phi$. The idea of the model is the following: suppose we know the threshold energy dependence of the differential cross section of a process $N_1N_2\rightarrow (N_{1^\prime}N_{2^\prime})_{BS} \, \phi$, while $(N_{1^\prime}N_{2^\prime})_{BS}$ denotes the bound state (BS) of two components $N_{1^\prime}$ and $N_{2^\prime}$ with binding energy $\varepsilon_B$. Then we can derive --- by analytical continuation of the final state wavefunction --- the energy dependence of the total cross section of a process $N_1N_2\rightarrow (N_{1^\prime}N_{2^\prime})_{SS} \, \phi$ with a scattering state $(N_{1^\prime}N_{2^\prime})_{SS}$ of same quantum numbers as the BS. 
According to F\"aldt and Wilkin the BS-wavefunction and the corresponding SS-wavefunction are related through an analytical continuation by $\lim\limits_{\kappa\rightarrow i \alpha} \{ \sqrt{2\,\alpha \,(\alpha^2 + \kappa^2)} \; \psi_{SS} (\kappa , r) \} = - \, \psi_{BS} (r)$ with $\alpha = \sqrt{\mu_{12} \, \varepsilon_B}$ and $\mu_{12}^{-1} = m^{-1}_{1^\prime} + m^{-1}_{2^\prime}$. As in a Watson-Migdal picture the T-matrix factorises into wavefunctions, there is now the hope, that also the T-matrix elements of the two processes are related by\footnote{The approximation made in the approach of G.\ F\"aldt and C.\ Wilkin is, that one has to match the T-matrix element of a $2\rightarrow 2$ process (described by 2 independent the Lorentz-invariants $s,t$) to a T-matrix element of a $2\rightarrow 3$ process (described by 5 independent the Lorentz-invariants $s,s_1,s_2,t_1,t_2$). This is only possible by imposing some approximative (onshell) constraints to $s,s_1,s_2,t_1$ and/or $t_2$.} \cite{fal97} ($q^\prime := |\vec{k}_\phi|_{cm}$):
\begin{equation} \frac{|T_{fi} (N_1N_2\rightarrow (N_{1^\prime}N_{2^\prime})_{SS}\,\phi\,)\,|^2}{(2\pi)^3 \, 2\, m_{1^\prime} \; (2\pi)^3 \, 2\, m_{2^\prime}} \simeq \frac{|T_{fi} (N_1N_2\rightarrow (N_{1^\prime}N_{2^\prime})_{BS}\,\phi\,)\,|^2}{2\,\alpha \,(\alpha^2 + \kappa^2) \; (2\pi)^3 \, 2\, m_{BS}} \, .\end{equation}  
$d\,\sigma (N_1N_2\rightarrow (N_{1^\prime}N_{2^\prime})_{BS}\,\phi)/d\Omega \propto \frac{q^{\,\prime}}{\kappa_0} \; |T_{fi} (N_1N_2\rightarrow (N_{1^\prime}N_{2^\prime})_{BS}\,\phi\,)\,|^2 \propto q^{\,\prime \,n}$ at threshold yields $|T_{fi} (N_1N_2\rightarrow (N_{1^\prime}N_{2^\prime})_{SS}\,\phi\,)\,|^2 \propto q^{\,\prime \,n-1}/(2\,\alpha \,(\alpha^2 + \kappa^2))$. After performing the phasespace integration one obtains for the total cross section with the $SS$-system in the final state ($\bar{\mu}^{-1}:=(m_{1^\prime}+m_{2^\prime})^{-1}+\, m^{-1}_\phi$):
\begin{equation} 
 \sigma (N_1N_2\rightarrow (N_{1^\prime}N_{2^\prime})_{SS}\,\phi) \; \propto \; \eta_\phi^{\,n} \; P^{\,(n)}(\eta_\phi \, \zeta_\phi)  \quad \mbox{with} \quad \zeta_\phi := \frac{m_\phi}{\sqrt{2\,\bar{\mu}\,\varepsilon_B}} \, .
\end{equation}
Some of the functions $P^{\,(n)}(\eta_\phi \, \zeta)$ are listed without and/or with Coulomb-corrections in \cite{fal97}. As mentioned in \cite{fal97,kle00b}  the cross section $\sigma (pp\rightarrow pp\pi^0)$ close to threshold can be well described by $\sigma (pp\rightarrow pp\pi^0) \propto \eta_\pi P^{\,(1)}(\eta_\pi \, \zeta_\pi) = \eta^4_\pi \, \zeta_\pi^3 /(4\,(1+\sqrt{1+\eta^2_\pi\zeta^2_\pi})^2)$ indicating a quasi-bound state in the corresponding $\{pn\}_{I=1}$ final state. For the discussion of the relation between the $\{pn\}_{I=0}$ system and the deuteron in the final state see e.g.\ \cite{fal97,uzi01,aba01}. The $N\Lambda K$-final state is addressed e.g.\ in \cite{sib98c}.
\subsection*{Onshell Faddeev models}
For completeness I want to mention here some works going in the direction of a full Faddeev calculation for $NN\rightarrow NNX$. Moalem et al. (1995) \cite{moa95b} developed a semiquantitative formalism for describing $T_{FSI}$ (and $T_{ISI}$) expanding the elastic 3-body T-matrix in the final state according to a Faddeev expansion upto second order.
First solutions of simplistic nonrelativistic Faddeev models using separable interaction kernels can be found for the $\eta d$-system in \cite{gar00,wyc01} and for the reaction $pp\rightarrow pp\eta$ in \cite{pen01}. The main result of the works is an investigation of the effective range parameters in the $\eta N$ system and the $\eta NN$ coupling constant. The models have to be improved to be quantitative and predictive\footnote{It e.g.\ should be remarked, that the complex $\eta N$ scattering length can't be investigated without considering also the effective range. For the description of a meson nucleon interaction potential parameters, which reproduce phaseshifts, have to be choosen consistently. Finally separable interaction kernels may introduce spurious states (also quasi-bound states) not existing in the original theory yet seen in the phaseshifts.}.
\section{Relativistic OWMA --- the  ABSSM}
As discussed in Section \ref{sec3} the Watson-Migdal Approach has been based on the idea, that the energy dependence of the T-matrix close to the reaction threshold is mainly driven by (``factorised'') destorted wavefunctions of interacting subsystems in the initial or final state, which are usually determined in nonrelativistic Schr\"odinger frameworks. 
With the exception of lightfront approaches the consistent factorisation of relativistic wavefunctions --- so called n-particle Bethe-Salpeter Amplitudes --- of bound and scattering states from the T-matrix has upto now been an open problem\footnote{In decay processes there is the general belief, that due to R.\ van Royen and \mbox{V.F.\ Weisskopf} \cite{roy67} the decay rate is --- to a good approximation --- proportional to the square of the wavefunction of the decaying system at its origin. Yet this assumes the factorisability of this square of the wavefunction.}. In order to remove this lack of formalism and taking into account the fact, that production processes of heavy mesonic systems close to threshold involve high energy and momentum transfers, the author recently developed the relativistic so called {\em Asymptotic Bethe Salpeter State Method} (ABSSM) \cite{kle99a,kle98,kle01b}, which has been applied \cite{kle99a,kle96} as a first test in a simplistic way to the reaction $pp\rightarrow p \,\Lambda \,K^+$ in the quark-gluon picture\footnote{It should be mentioned that publication \cite{kle96} contains apart from a quark-gluon description of $pp\rightarrow p\,\Lambda K^+$ the first theoretical investigation of this reaction in the nucleon-meson picture {\em taking into account} beside the pseudoscalar meson-exchanges ($\pi,K$) the upto now by different authors neglected, yet --- as shown in \cite{kle96} --- {\em important vector-meson exchanges ($\rho,K^\ast$)}.}. In a second step the ABSSM is now applied to reactions like $NN\rightarrow d\,X$. How is the relativistic deuteron wavefunction factorised within the ABSSM? A composite state vector $|P,B>$ being normalised according to $\ll P^\prime,B^\prime|P,B>=(2\pi)^{\,3} \, 2 \,\omega_B(|\vec{P}\,|) \, \delta^{\,3} (\vec{P}^{\,\prime} - \vec{P}) \, \delta_{B^\prime B}$ can be expanded into free Fock states. The overlap of the free n-particle Fock states with $|P,B>$ is then reexpressed in terms of respective n-particle Bethe-Salpeter Amplitudes. For details I refer to \cite{kle01b}. The most dominant term of the composite deuteron state vector is the two nucleon sector, i.e.\footnote{Here $s_i,t_i$ denote spin and isospin. Spinors/operators fulfil $\bar{u} \, (\vec{p},s,t) \,u \, (\vec{p},s^\prime,t^\prime) = 2\, m \, \delta_{ss^\prime} \, \delta_{t t^\prime}$, $\{ b \, (\vec{p},s,t)\, , \, b^+ \, (\vec{p}^{\,\prime},s^{\,\prime},t^{\,\prime})\}= (2\pi)^3 \, 2 \,\omega(|\vec{p}\,|) \, \delta^{\,3} (\vec{p} - \vec{p}^{\,\prime}) \, \delta_{s s^\prime} \, \delta_{t t^\prime}$, $\ldots$}:
\begin{eqnarray}
\lefteqn{|d^+(P,M)> \quad \simeq} \nonumber \\
 & \simeq &
\sum\limits_{s_1,t_1} \sum\limits_{s_2,t_2}
\int 
\frac{d^3p_1}{(2\pi )^3 \, 2 \,m_{{}_N}} 
\frac{d^3p_2}{(2\pi )^3 \, 2 \,m_{{}_N}}
\int d^3x_1 \; d^3x_2 \;\;
e^{ \displaystyle -i\, (\vec{p}_1 \cdot \vec{x}_1 + 
 \vec{p}_2 \cdot \vec{x}_2 )} 
\nonumber \\
& & 
\qquad\quad b^+ \, (\vec{p}_1,s_1,t_1) \; b^+ \, (\vec{p}_2,s_2,t_2) |0> \; 
\bar{u}^{(1)} \, (\vec{p}_1,s_1,t_1) \;
\bar{u}^{(2)} \, (\vec{p}_2,s_2,t_2)
\nonumber \\ 
 & & 
\qquad\quad \ll 0| \, T \, [ \, \psi^{(2)}  (x_2) \; \psi^{(1)}  (x_1) \, ] \, |d^+(P,M)> 
 \, \Big|_{x^0_1=x^0_2=0} \; .
\end{eqnarray}
In this form the composite state vector can be conveniently used for the calculation of T-matrix elements with the help of Wick's Theorem in the standard manner. In phasespace integrations  the composite deuteron appears only as one particle with mass $m_d$. It is beyond the scope of this presentation to discuss all the promising results and relativistic corrections predicted by this new method\footnote{A further decisive quantitative test of the ABSSM will be its application to the positronium decay compared to standard Bethe-Salpeter descriptions or the new factorisation method based on a dispersion approach presented in \cite{pes01}.}.
\section{What can go wrong?}
\subsubsection*{The unitarity violation problem}
Naive use of enhancement factors may violate unitarity, especially in many channel problems! $T_{FSI}$ and $T_{ISI}$ are related by unitarity!
\subsubsection*{The fitting problem}
Accurate experimental measurements of the total cross section of $NN\rightarrow NNX$ close to threshold are supposed to prove or disprove theoretical descriptions of the meson nucleon dynamics. As is seems, that the energy dependence of total cross sections close to threshold is mainly described by FSI parametrised e.g. by a sophisticated Effective Range Theory taking into account eventual quasi-bound states, the experimental measurement of {\em total} cross section constrain essentially one number, i.e.\ $|T^{(0)} \, T_{ISI}|$. Under the assumption, that we can estimate $|T_{ISI}|$ quantitatively, we can constrain the modulous of the short-ranged T-matrix $|T^{(0)}|$.  It is therefore curious to observe, that the majority of present theoretical models for the description of $NN\rightarrow NNX$ introduce or keep one free parameter in their models, which allows them to bring their calculated cross sections into the experimental data. The most distributed method (see e.g.\ \cite{nis98,ber99,mil91}, the seventh equation in \cite{nis98} and the second equation in \cite{han99}) of introducing a ``fitting parameter'' is the following: As people know from the WMA, that $T_{FSI}$ is proportional to the dimensional quantity $\psi_f (r=0) \sim 1/(\kappa (\cot\delta - i))$, they introduce a dimensional constant or function $N$ (``numerator'') to obtain a dimensionless enhancement factor $T_{FSI}$ by $T_{FSI} = N/(\kappa (\cot\delta - i))= N/(-\, a^{-1} + r\,\kappa^2/2 + \ldots - i\, \kappa)$. This numerator $N$ is either fitted to the data or chosen without much reasoning\footnote{As it has been shown in Section \ref{sec3}, the numerator is given by $N = \kappa \cot\delta - i\, \kappa P (\kappa) = -\,1/a^{-1} + r\,\kappa^2/2 + \ldots - i\, \kappa \, P(\kappa)$ with $P(\kappa)$ staying undetermined. Even in the OWMA, i.e. with $P(\kappa)\simeq 0$, there is no reason to neglect in $N$ the term $r\,\kappa^2/2$, while keeping it finite in the denominator. Such a procedure strongly affects the energy dependence of the cross section.} to be $ -\,a^{-1}$.
It is also possible to transfer the quantitative uncertainty in the description of FSI/ISI to a ``fitting parameter'' in the short ranged T-matrix $T^{(0)}$. 
While in the model of V.\ Bernard et al.\ \cite{ber99} $|T^{(0)}|^2$ is just chosen to fit to the data,
the ``fitting parameter'' of the J\"ulich-model for the threshold production of pions in nucleon nucleon collisions (see \cite{hai00} and references therein) is the heavy meson nucleon (anti-)nucleon coupling constants ($\omega NN$, $\sigma NN$) in the so called ``pair diagrams'', which is tuned to meet the $pp\rightarrow pp\pi^0$ total cross section. 
\subsubsection*{The Gell-Mann-Watson-Rosenfeld-Koltun-Reitan et al.\ problem}
The interpretation of the total cross section of $pp\rightarrow pp\pi^0$ close to threshold has a peculiar history. The the theoretical situation is summarised by H.\ Machner and J.\ Haidenbauer (1999) \cite{mac99} as follows: ``From phase space alone the cross section should follow a $\ldots \; \eta^4$ dependence. However, this is in complete disagreement with data. Taking into account `minimal effects' from the final NN interaction i.e.\ using the first term in the ERE\footnote{At this point the authors refer to the expression $|T_{FSI}|^2 = 1/ (k^2 + [-1/a + r\,k^2/2]^2)$.} $\ldots$, leads to an $\eta^2$ dependence $\ldots$.''. Then the authors give a widely and by now used expansion of $\sigma (pp\rightarrow pp\pi^0)$ close to threshold, which they call ``barrier penetration model'', i.e.
\begin{equation}\sigma (pp\rightarrow pp\pi^0) \stackrel{?}{=} \beta_{Ss} \, \eta_\pi^{\,2} + \beta_{Ps} \, \eta_\pi^{\,6} + \beta_{Pp} \, \eta_\pi^{\,8}\, .\end{equation} 
This threshold expansion goes back to a phenomenological discussion of M.\ Gell-Mann, K.M.\ Watson (1954) \cite{gel54} and A.H.\ Rosenfeld (1954) \cite{ros54}. Early datapoints for $\eta_\pi \ge 0.5$ listed in \cite{sta58} seemed to support the $\eta_\pi^2$-behaviour of $\sigma (pp\rightarrow pp\pi^0)$ very close to threshold, while the 2 datapoints at $\eta_\pi=0.660$ and $1.11$ \cite{ros54} indicated $\sigma (pp\rightarrow pp\pi^0) \propto \eta^{\,8}_\pi$ (see also \cite{fer55}) showing the onset of $Pp$-production. This is why theory thankfully acknowledged the theoretical model for $\sigma(pp\rightarrow pp\pi^0)$ presented by D.S.\ Koltun and A.\ Reitan (KR1966) \cite{kol66}, which resulted in $\sigma(pp\rightarrow pp\pi^0) \, \propto \, \eta^{\,2}_\pi$ close to threshold\footnote{It should be mentioned, that 1969 there followed a complementary theoretical approach by M.E.\ Schillaci et al.\ \cite{sch69}.}. Accurate data for $0.13 \le \eta_\pi \le 0.56$ published by H.O.\ Meyer et al.\ (1990) \cite{mey90,mey92} and A.\ Bondar et al.\ (1995) \cite{bon95} could be --- within the errorbars --- understood as a final confirmation of the $\eta_\pi^2$-dependence of $\sigma (pp\rightarrow pp\pi^0)$ close to threshold\footnote{If one has a more careful look to the data, one can see already in these data very close to threshold a deviation of the energy dependence of $\sigma(pp\rightarrow pp\pi^0)$ from the ``expected'' $\eta_\pi^2$-behaviour.}.
Yet latest data (see \cite{zlo98, joh00} and references therein) show clearly, that the energy dependence of $\sigma (pp\rightarrow pp\pi^0)$ very close to threshold is much steeper than $\eta_\pi^2$. It seems, that phasespace behaviour, i.e.\ a $\eta_\pi^4$-dependence, is restored for $\eta_\pi < 0.2$, while the ``$\eta_\pi^2$-plateau'' of $\sigma (pp\rightarrow pp\pi^0)/\eta_\pi^2$ in the range $0.2\le \eta_\pi \le 0.5$ is a structure due to a quasi-bound state in the $pp$-system well described within the formalism of G.\ F\"aldt and C.\ Wilkin \cite{fal96a,fal96b,fal96c,fal97} (see also Section \ref{sec5}). It is no surprise, that --- according to the introduction of \cite{dae01} --- the most theoretical (standard) models of $pp\rightarrow pp\pi^0$ are more or less sophisticated versions of the work of KR1966, who differ mainly in the sophistication of $T^{(0)}$, yet use same FSI and phasespace arguments in order to reproduce the $\eta^2_\pi$-dependence of $\sigma (pp\rightarrow pp\pi^0)$ close to threshold. The problem is now, that the calculations of KR1966 and therefore also the --- often quoted --- original considerations of M.\ Gell-Mann, K.M.\ Watson (1954) \cite{gel54} and A.H.\ Rosenfeld (1954) \cite{ros54} contain an explicit mathematical mistake. To make the mistake more transparent, I translate the notation of KR1966 in the language used in this presentation, i.e. I perform the replacements $M\rightarrow m_p$, $\mu \rightarrow m_\pi$, $p^\prime \rightarrow \kappa$,  $E_p \rightarrow T^{\,\prime} := T_{{pp}^\prime}^{\,kin} = \sqrt{s_1} - 2\, m_p$ and $E_f \rightarrow Q_{cm}$. Using this notation equations (18) and (19) in KR1966 read:
\begin{equation} \frac{d\sigma}{dT^{\,\prime}} = \frac{2\,\pi}{v}\;  \frac{1}{4} \; \sum |T_b|^2 \; \frac{d\rho\,(Q_{cm})}{dT^{\,\prime}} , \; \frac{d\rho\,(Q_{cm})}{dT^{\,\prime}} = \frac{(2\,m_\pi\,m_p)^{\frac{3}{2}}}{(2\pi)^4} \sqrt{(Q_{cm} - T^{\,\prime})\, T^{\,\prime}} .
\end{equation}
Making according to KR1966 the approximation $|T_b|^2 \simeq |T(0)|^2/(1+a^2 \, \kappa^2)$ and using $\kappa^2 = m_p \, T^{\,\prime} + O (T^{\,\prime \, 2})$ we obtain
\begin{equation} \sigma \simeq \frac{2\,\pi}{4\,v} \sum |T(0)|^2 \int^{\,Q_{cm}}_0 dT^{\,\prime} \; \frac{1}{1 + a^2 \, m_p \, T^{\,\prime} + O (T^{\,\prime \, 2})} \; \frac{d\rho\,(Q_{cm})}{dT^{\,\prime}} \propto Q^2_{cm} \propto \eta^4_\pi .
\end{equation}
This should be compared to equation (21) in KR1966. Koltun and Reitan write: ``For $Q_{cm}\gg m^{-1}_p \, a^{-2}$ we may approximate Eq. (18) and write'':
\begin{equation} \sigma \simeq \frac{2\,\pi}{4\,v} \frac{1}{m_p\,a^2} \sum |T(0)|^2 \int^{\,Q_{cm}}_0 dT^{\,\prime} \; \frac{1}{T^{\,\prime}} \; \frac{d\rho\,(Q_{cm})}{dT^{\,\prime}} \; \propto Q_{cm} \propto \eta^2_\pi .
\end{equation}
i.e.\ they obtain a total cross section with a $\eta_\pi^2$- and not a $\eta_\pi^4$-dependence. We can see, how the error occurs: although the lower integration limit is 0, they allow the replacement $1 + a^2 \, m_p \, T^{\,\prime}\rightarrow a^2 \, m_p \, T^{\,\prime}$, which is equivalent neglecting the $- a^{-1}$ term in the ERE\footnote{Even a infinitesimal value for $1/a$ would yield $\sigma(pp\rightarrow pp\pi^0)\propto\eta^4_\pi$ close to threshold!}. As easy the error is detectable in KR1966, as hidden it is among various features of $T^{(0)}$ in the publications and computer codes of present theoretical calculations. One either can recover it by trying to reproduce their results or from the statements they make in their publications.
Many models mentioned in E.\ Hern\'{a}ndez et al. \cite{her95,her99}, especially the models of C.J.\ Horowitz et al.\ \cite{hor94}, J.A.\ Niskanen \cite{nis92,nis96}, G.A.\ Miller et al. \cite{mil91} and the J\"ulich-model (see \cite{hai00} and refererences therein), are set up nearly in the same way as KR1966. As all these works seem to reproduce the energy dependence of the direct pion production close to threshold predicted in Koltun and Reitan \cite{kol66}, it seems, that they all use the same approximation as Koltun and Reitan \cite{kol66}\footnote{This is quite surprising, as J.\ Niskanen states in \cite{nis92}, that H.O.\ Meyer (see e.g.\ p.\ 657 in \cite{mey92}), G.A.\ Miller, P.U.\ Sauer \cite{mil91} are aware, that the cross section ``dependence at threshold is not $\sigma_{tot} = a \, \eta^2$ \ldots''. In their case this seems to be due to the energy dependence of their specific form of $T^{(0)}$ (see e.g.\ (\ref{lab49})) and not due to a corrected phasespace.}. 
Ch.\ Hanhart states in Tab.\ 1.2 of his doctoral thesis \cite{han98}, that $\sigma_{11} \propto \eta^2$ and on page 13 below eq.\ (1.7), that $\sigma^{(0,\ell)}_{NN\rightarrow NN\pi} (\eta) \propto \eta^{2\ell + 2}$ --- in contrast to my result (\ref{eqx1}) ---, even after having derived the correct integral (\ref{eqx2}).
\mbox{N.\ Kaiser} claimed \cite{kai99}, that the $1/a$ term in the ERE could be neglected, from wich they obtained in \cite{ber99} an $Q_{cm}$-dependence of the cross section close to threshold.
An alternative explanation for the mysterious threshold energy dependence of modern calculations hidden in computer codes might be another systematic mistake, i.e.\ the use of a 2-body phasespace instead of a 3-body phasespace. E.g. J.A.\ Niskanen states in \cite{nis98}, that $\sigma \sim (a^2 \,p_f)/(1 + (p_f \, a)^2)$, where ``$p_f$ in the numerator is the momentum dependence of the phase space''. Well, from the arguments given above we know, that the momentum dependence must be $p^2_f$! 
Finally I want to mention here a semantic problem in the definition of $\eta_\pi$: in \cite{mey90} H.O.\ Meyer et al.\ state correctly, that $\eta_\pi$ is ``the largest possible center-of-mass pion momentum (with nucleons at rest relative to each other) divided by the pion mass'', which yields $\eta_\pi := \sqrt{\lambda (s,m^2_\pi,s^{\,min}_1)/(4\, s\,m^2_\pi)}$. In Horowitz et al. (1994) \cite{hor94} it is stated, that $q^\prime$ is not $\kappa$, but ``the pion momentum in the final $pp$ center-of-mass system'', which would yield the definition  $\eta^{Horowitz}_\pi := \sqrt{\lambda (s,m^2_\pi,s^{\,min}_1)/(4\, s^{min}_1\,m^2_\pi)} \stackrel{!}{=} \eta_\pi \, \sqrt{s/(2\,m_p)^2}$. Even if both quantities vanish at threshold, they increasingly differ at higher energies. $\eta_\pi=0.557$ yields e.g. $\eta^{Horowitz}_\pi=0.603$.
\subsubsection*{The Miller-Sauer problem}
The energy dependence of the total cross section for $pp\rightarrow pp\pi^0$ derived by G.A.\ Miller and P.U.\ Sauer (1991) \cite{mil91} --- the standard calculation for direct meson production --- is not reproducable by means discussed in their publication. The fall-off of the total cross section for $\eta_\pi > 0.4$ is much too steep compared to Effective Range or Coulomb Modified Effective Range calculations or for a result obtained by a calculation with wavefunctions as solutions of Schr\"odinger equations with strong and Coulomb potential. The direct meson production is only possible through FSI and ISI and is therefore a crucial test for a correct description of FSI and ISI.
\subsubsection*{The problem of wrong application of the two potential formalism}
Many people nowadays describe the influence of Coulomb interactions in FSI by the CFS-CMERE \cite{cin1,noy72} listed in (\ref{eqcfs}) and derived from a two potential formalism. I.e. the total S-wave phaseshift $\delta_0(\kappa)$ is given by $\delta_0 (\kappa) = \sigma_0 (\kappa) + \delta_0^{\,c} (\kappa)$, while $\sigma_0 (\kappa)$ is the S-wave Coulomb phaseshift. 
In the construction of $T_{FSI}$ we have --- according to the two potential formalism --- to use $\delta_0 (\kappa)$ and not $\delta_0^{\,c} (\kappa)$. Therefore $T_{FSI}$ is given in the OWMA according to (\ref{eqx5}) by
\begin{eqnarray} \lefteqn{T_{FSI} \quad \simeq} \nonumber \\
 & \simeq & \frac{1}{2} \left( e^{\,2\,i\,\sigma_0 (\kappa)} \, \frac{- \; \frac{1}{a^c} \; + \; \frac{1}{2} \; r^c \; \kappa^2 - \frac{p\;\kappa^{\,4}}{1 + q\; \kappa^2}  - 2 \, \kappa \, \eta \, H(\eta) + \, i \,\kappa \; {\cal C}_0^{\,2} (\eta )}{- \; \frac{1}{a^c} \; + \; \frac{1}{2} \; r^c \; \kappa^2 - \frac{p\;\kappa^{\,4}}{1 + q\; \kappa^2}  - 2 \, \kappa \, \eta \, H(\eta) - \, i \,\kappa \; {\cal C}_0^{\,2} (\eta )} + 1 \right) \, .\nonumber \\
 \end{eqnarray}
The limit $\sigma_0 (\kappa) \rightarrow 0$ performed by the majority of experimental and theoretical researchers yields results, which one would get, {\em if one does not apply the two potential formalism properly}. As discussed above the presence of Coulomb phaseshifts $\sigma_0 (\kappa)$ introduces essential singularities in the phasespace integrations, when calculating cross sections, which can only be resolved, if the Coulomb potential is regularised consistently.
\subsubsection*{The numerator or Moalem et al.\ problem}
Going back to the ideas of K.\ Brueckner et al. \cite{bru51}, i.e. that $T_{FSI} \propto \psi (r=0)$ or equivalently $T_{FSI} \simeq (N \exp (i\,\delta) \sin \delta) / \kappa = N/(\kappa \, (\cot\delta - i))$, it has been very early clear, that the denominator of enhancement factors is determined by the ERE, i.e. by $ \kappa \, (\cot\delta - i) = - a^{-1} + \ldots - i\,\kappa $. The unknown numerator of $T_{FSI}$ has been fixed either by hand, i.e.\ by setting the dimensional constant $N$ to some suitable value, or --- to be free of ambiguities --- $T_{FSI}$ was chosen to be the inverse Jost function\footnote{This choice can be found in nearly every textbook of scattering theory (see e.g.\ \cite{gol64,joa83}).}, i.e. $T_{FSI} = 1/f_0(\kappa)$. Even if the inverse Jost function yields --- apart from the correct denominator of $T_{FSI}$ --- a $\kappa$-dependent numerator, it can be observed in literature, that many authors approximate the numerator of $T_{FSI}$ (after performing a Taylor expansion in $\kappa$) by the term, which is $O(\kappa^0)$, i.e.\ by a constant, while keeping the full $\kappa$-dependence of the denominator (see e.g.\ \cite{han99,sib98,sib98b,sib01})\footnote{It should be noted, that this nontrivial approximation changes the energy dependence of the total cross section at larger energies.}.

Among others A.\ Moalem et al. (1996) \cite{moa96,moa95} and R.\ Shyam (1999) \cite{shy99} use within an Effective Range Approach complex scattering length and effective range parameters, Moalem for the $\eta p$ system in the final state, Shyam for the $K^+\Lambda$ system in the final state. Both authors use a Coulomb modified CFS formula \cite{cin1,noy72} to describe Coulomb interactions and {\em both authors observe mysterious ``oscillations'' of the cross section as a function of energy close to threshold}. 
How can this be, especially as both use different methods for enhancement factors: Moalem uses $T_{FSI} = N / \kappa (\cot \delta_0 (\kappa) - i)$, while Shyam uses the inverse Jost function $T_{FSI} = 1 / f_0 (\kappa)$. {\em The problem is, that both authors don't use the correct numerator}\footnote{By their choice both authors set the numerator of the enhancement factor to a constant: Moalem replaces $\kappa \cot \delta_0 (\kappa)$ by a constant, Shyam performs first the replacement $\mbox{Re} f_0(\kappa)\rightarrow 1$, then he removes the $\kappa^2$ term of the ERE in the numerator ending up with an enhancement factor like Moalem, i.e.\ $T_{FSI} = N^\prime / \kappa (\cot \delta_0 (\kappa) - i)$.}{\em in the enhancement factor}, which is determined by the OWMA, i.e. they must use $T_{FSI} = \kappa \cot \delta_0 (\kappa)/(\kappa ( \cot \delta_0 (\kappa) - i)) =  (\mbox{Re} f_0(\kappa)) /f_0(\kappa)$,
which stabilizes the energy dependence very close to threshold. {\em Both authors don't apply the two potential formalism correctly} by skipping the Coulomb-phaseshifts and the Rutherford amplitude. On the other hand one can observe, that --- due to the different numerators of $T_{FSI}$ --- the penetration factor ${\cal C}_0^2 (\eta )$ enters differently in the ``more correct'' (above) and wrong (below) expression:
\begin{eqnarray} \frac{\kappa \cot \delta^c_0 (\kappa)}{\kappa \, ( \cot \delta^c_0 (\kappa) - i)} & \rightarrow & \frac{- \; \frac{1}{a^c} \; + \; \frac{1}{2} \; r^c \; \kappa^2 + \ldots - 2 \, \kappa \, \eta \, H(\eta) \qquad\qquad\;\quad}{- \; \frac{1}{a^c} \; + \; \frac{1}{2} \; r^c \; \kappa^2 + \ldots - 2 \, \kappa \, \eta \, H(\eta) - \, i \,\kappa \; {\cal C}_0^{\,2} (\eta ) } \nonumber \\ 
\frac{N}{\kappa \, ( \cot \delta^c_0 (\kappa) - i)} & \rightarrow & \frac{N \; {\cal C}_0^{\,2} (\eta )}{- \; \frac{1}{a^c} \; + \; \frac{1}{2} \; r^c \; \kappa^2 + \ldots  - 2 \, \kappa \, \eta \, H(\eta) - \, i \,\kappa \; {\cal C}_0^{\,2} (\eta ) } \, . \qquad
\end{eqnarray} 
\subsubsection*{The Titov et al.\ problem}
As a side effect of the purpose of their calculation Titov et al.\ (2000) \cite{tit00} reconsidered to some extend the Inverse Jost Function Method for the description of ISI and FSI. The problem arising in their work is, that they use for the description of the $pp$ FSI the effective range parameters $a_{pp} = - 7.8098$ fm and $r_{pp} = 2.767$ fm, even not using a CMERE. In order to describe their inverse Jost function they  essentially use a simple nuclear ``shape independent'' ERE $\kappa \cot \delta (\kappa) = - a^{-1} + r\,\kappa^2 / 2$. In such a nuclear ERE without Coulomb they should use for the $pp$ system the well known effective range parameters $a_{pp} = - 17.1$ fm and $r_{pp} \simeq 2.8$ fm (see e.g. p.\ 254 in \cite{nag1}, p.\ 425 in \cite{rei68}, p.\ 115 in \cite{squ1}). The relation beween nuclear and Coulomb modified effective range parameters is discussed e.g.\ in \cite{hel1} or on p.\ 39 in \cite{lan2}.
\subsubsection*{The wavenumber problem}
The threshold production of heavy mesons in NN collisions involves high relative energies in the initial state and high energy- and momentum-transfers at the meson-production vertex. In quantitative calculations it is therefore important to define quantities carefully and such, that they are valid also at high energies within a relativistic framework. This is, why it is important to mention, that standard EREs are expansions in the Lab-wavenumber and not in momenta\footnote{It is therefore not correct, if on p.\ R241 in \cite{mac99} it is stated, that ``$k$ is the relative baryon momentum''. The authors should have written, that $k$ is the relative Lab-wavenumber, which makes of course only at high relative energies a difference! On page R242 in \cite{mac99} the authors define the Gamov factor by $C(q) = 2\,\pi \, \gamma_q /(\exp(2\,\pi\,\gamma_q) - 1)$ with $\gamma_q = \mu\alpha /q$ and state, that ``$\mu$ and $q$ are the reduced mass and the relative momentum of the two-body subsystem in which the Coulomb interaction occurs $\ldots$''. This definition is also not exact at higher energies! They should have defined $\gamma_q = \alpha /v$ with $v$ being the relative Lab-velocity of the two-body subsystem in which the Coulomb interaction occurs.}.
\subsubsection*{The phaseshift problem}
Phaseshifts are described in the Lab-system! This is not only true for proton nucleus scattering, but also for $pp$ scattering, where the cm-system seems more convenient. If phaseshifts are theoretically determined from a Schr\"odinger equation (e.g.\ for the $pp$-system (p.\ 419ff in \cite{rei68}), it makes a large difference at higher energies, whether one uses Lab- or cm-quantities:
\begin{eqnarray}  & & \left( \frac{1}{r^2} \frac{d}{dr} \left( r^2 \frac{d}{dr} \right) 
+ K^2 - \frac{L(L+1)}{r^2} - \frac{\alpha \, m_p}{r} \right) \; 
R^{\, J,I}_{L,S} (K ,r) \quad = \nonumber \\
 & & \qquad\qquad\qquad\qquad\qquad\qquad = \quad  
m_p \, \sum\limits_{L^\prime S^\prime} 
 V^{J,I}_{LS, L^\prime S^\prime} (\tilde{r}) \; R^{\, J,I}_{L^\prime S^\prime} (K ,r) \, .
\qquad \end{eqnarray}
($\tilde{r}_{cm} = r$, $\tilde{r}_{\,\mbox{\scriptsize Lab}} = r/2$). I.e.\ use Lab-quantities $K=\kappa$ and $T_{\,\mbox{\scriptsize Lab}}= \omega_p (2\kappa)-m_p 
\simeq 2 \, \kappa^2 \, / m_p$ (Final State (FS)) or
$K=\kappa_{\,0}$ and $T_{\,\mbox{\scriptsize Lab}}= 
\omega_p (2\kappa_{\,0})-m_p \simeq 2 \, \kappa_0^2 \, / m_p$ (Initial State (IS)) and {\em not} the cm-quantities $K=2m_p \, \kappa/\sqrt{s_1}$ and $T=\sqrt{s_1}-2m_p$ (FS) or $K=2m_p \, \kappa_{0}/\sqrt{s}$ and $T=\sqrt{s}-2m_p$ (IS)!
\subsubsection*{The transition operator problem}
In traditional threshold production models for direct $\pi^0$-production in $pp\rightarrow pp\pi^0$ close to threshold people don't use one consistent $\pi^0$-production operator. There are e.g.: 
\begin{eqnarray} T_{fi} & \propto &
 \int\limits_0^\infty dr \; r^2 \;  
(R^{\, 01}_{00} (\kappa ,r))^\ast \; j_{0} (\kappa_{\pi^0} \, r) \;
\left( \, \frac{d}{dr} + \frac{2}{r} \, \right) 
R^{\,01}_{11} (\kappa_{\, 0} ,r) \; \mbox{$\rightarrow$\cite{kol66,mey92,nis96}} \nonumber \\
T_{fi} & \propto & 
 \int\limits_0^\infty dr \; r^2 \;  
(R^{\, 01}_{00} (\kappa ,r))^\ast \; j_{0} (\kappa_{\pi^0} \, r) \;
\left( \, \frac{d}{dr} + \frac{1}{r} \, \right) 
R^{\,01}_{11} (\kappa_{\, 0} ,r)\; \mbox{$\rightarrow$\cite{hor94}} \nonumber \\
T_{fi} & \propto &
 \kappa_{\pi^0}  
\; \int\limits_0^\infty dr \; r^2 \;  
(R^{\, 01}_{00} (\kappa ,r))^\ast \; j_{1} (\kappa_{\pi^0} \, r) \;
R^{\, 01}_{11} (\kappa_{\, 0} ,r) \;\;\; \mbox{$\rightarrow$\cite{nis98}} \, . \label{lab49}
\end{eqnarray}
which are more or less related by a partial integration\footnote{It is unclear, why various terms appearing during this partial integration should vanish or be neglected. The transition operator describes mainly the FSI in the NN-system, as $j_0(\kappa_{\pi^0} \, r)$ representing the produced $\pi^0$-meson is nearly constant close to threshold.}. For a correct description of FSI we need the correct transition operator.
\subsubsection*{The data reduction problem}
For a FSI reduction of their data experimentalists tend to multiply enhancement factors describing the FSI of possible particle subsystems in the final state. I quote here e.g.\ \mbox{J.\ Smyrski} et al.\ \cite{smy00}, who states: ``The $pp\eta$ FSI can be factorized into $pp (f_{pp})$ and $p\eta (f_{p\eta})$ factors and integrated over the available phase-space volume $\rho_3$: $\sigma (\varepsilon) \sim \int f_{pp} (q_{pp}) \cdot f_{p\eta} (q_{p_1\eta}) \cdot f_{p\eta} (q_{p_2\eta}) \, d\rho_3$, $\ldots$''. {\em For the moment there is no obvious theoretical reason to choose a product of final state factors rather than a sum or anything else to describe the FSI!} Well, the leading terms of a Faddeev expansion would suggest a the coherent square of a sum. It is an important theoretical issue in the future, to give a satisfactory answer to this still open problem.
\section{Conclusions and outlook}
\subsection*{Theoretical aspects}
We come to the conclusion, that --- in order to be able to relate quantitatively and conclusively theoretical quantities to experimental data --- one either has has to calculate the process $NN\rightarrow NNX$ close to threshold completely, e.g.\ within a three-body Faddeev approach, or there is a {\em need} of a quantitative theory to separate and estimate Initial-/Final-State effects. It is unsatisfactory and demanding, that --- in spite of the vast related activities of theoreticians during the passed about 60 years --- {\em such a theory for estimating Initial-/Final-State effects does not exist yet!} This theory also has to be extended to differential cross sections for processes involving unpolarized and/or polarized particles close to threshold to extract independent information from data. Even being under control in the differential description of elastic scattering of charged particles the application of the theory of Coulomb scattering in initial and final states of particle number changing reactions is --- due to the infinite range of the Coulomb interaction and the offshellness of the particles in the interaction zone --- a highly nontrivial problem, requiring well defined regularization schemes, especially close to the particle production threshold, at which produced charged particles have a long time to interact. The development of an adequate theoretical --- if possible, relativistic --- formalism is an outstanding task. In order to handle interference effects and inelasticities a consistent field theoretical treatment of intermediate resonant states is demanded. An adequate formalism for the treatment of Fermionic \cite{kle99a,kle99b,kle00} and Bosonic \cite{kle01a} resonances has been recently proposed. Keeping in mind the high relative energies and momenta of the incoming nucleons and the high momentum transfers involved in production processes of heavy mesons, relativistic formulations for nucleon induced meson production close to threshold are crucial. This includes the consideration of boost effects and e.g.\ tensor components, negative energy and parity components of the deuteron wavefunction within quantitative calculations. For the consistent relativistic treatment of bound of quasi-bound composite systems in the initial and final state the author refers to the new available formalism called ABSSM \cite{kle01b}. It is a pitty to observe, that the majority of theoretical models stays nonrelativistic, especially because most of the traditional models have been developed to describe pion production only.
Let's go for solving relativistic Faddeev problems! For the $NN\pi$-system I refer e.g. to \cite{kvi98,phi96}.
Not everything what is written should be believed!
\subsection*{Experimental aspects}
Thanks to  experimental facilities like the COSY in J\"ulich, the CELSIUS ring in Uppsala or the IUCF in Bloomington the accuracy of cross section data related to nucleon induced meson production processes close to threshold improved significantly within the last ten years. It is a challenging, yet unexpected difficult task for theory to handle the experimental results {\em quantitatively}. In order to obtain as much independent experimental information as possible to constrain theory it is {\em highly desirable} to continue with the experimental investigation of differential cross sections, polarization and polarization transfer observables, production of heavy mesonic systems with and without strangeness, processes involving a deuteron beam or target and/or further light and heavy nuclei. For a quantitative estimate of the effect of ISI on nucleon induced meson production reactions close to threshold an experimental determination of $NN$ scattering phaseshifts and inelasticities at high relative energies is crucial. As proposed in \cite{mos01} it is highly desirable to compare experimentally uncharged and charged reaction channels like e.g.\ $nn\rightarrow nnX$ and $pp\rightarrow ppX$ to learn something about the influence of Coulomb effects in the Initial/Final State and the effective range parameters of uncharged strong interacting particle systems. In the meantime we ask for being patient with slowly theoreticians!\footnote{{\bf Acknowledgements:} F.K.\ thankfully acknowledges the kind invitation to the COSY-11 collaboration meeting and the long-term interest in and support of the presented work by K. Kilian and M.\ Dillig. For young physicists like the author it is very hard to develop independent research in an extensive field of established theoretical phenomenology, if there were not people, who believe, that theory should be predictive and quantitatively constrained by experiment.
This work dedicated to my friend and experimental collegue \mbox{P.\ Moskal} has been supported mainly by the Forschungszentrum J\"ulich under contract No.\ ER-41154523,  COSY FFE-project No.\ 41324880 and in part by the 
{\em Funda\c{c}\~{a}o para
a Ci\^{e}ncia e a Tecnologia} \/(FCT) of the {\em Minist\'{e}rio da
Ci\^{e}ncia e da Tecnologia} \/of Portugal, under Grant no.\ PRAXIS
XXI/BPD/20186/99.} 
{
}

\end{document}